# Constitutive theory of saturated porous media considering porosity-dependent skeleton strain and chemical activity

Ya-yuan Hu [a*], Shu-hang Yuan [a]

(*Corresponding author)

[a] Research Center of Coastal and Urban Geotechnical Engineering, Zhejiang University, Hangzhou 310058, China.

**Abstract**: In order to reveal the coupling effect among the chemical activity and the hydraulic seepage as well as the mechanical properties, a constitutive theoretical framework considering the chemical activity for saturated porous media is derived from the mixture theory incorporated with the chemical thermodynamics. First, to highlight the important role of porosity in the hydro-mechanical-chemical multi-field coupling mechanism, the solid strain is divided into the porosity-dependent skeleton strain, the matrix strain and the mass-exchange strain. The stress and strain state variables are determined from the energy-conjugated form of the energy balance equation for establishing constitutive equations. Second, under infinitesimal case, general elastic constitutive equations including the relationship between mass fraction and its chemical potential are expressed by the free energy potential. Plastic model and the constitutive equation of thermodynamic flux and force are derived from the dissipative potential. Finally, under the guide of this theoretical framework, the complete swelling constitutive models in the confined compression are established for bentonite. The corresponding governing equations are formulated for multi-field two-phase saturated porous media. Compared with the experimental data, the proposed model can well reflect the chemical and mechanic coupling characteristics of representative elementary volume in the different NaCl concentrations of solution for saturated bentonite.

**Key words** Saturated porous media; hybrid mixture theory; mass fraction; Chemical potential; Skeleton strain; Constitutive relationship;

**Nomenclature**

$\alpha$   indicator variable of phase $\alpha \in \{S,F\}$

$\beta$   indicator variable of constituent in the fluid phase $\beta \in \{L,c\}$

$\varphi_S$, $\varphi_F$   volume fractions of the solid phase and fluid phase.

$\varphi_{S0}$, $\varphi_{F0}$   initial volume fractions of the solid phase and fluid phase.

$\rho_S$, $\rho_F$   average densities of the solid and fluid phases.

$\rho_{S0}$, $\rho_{F0}$   initial average densities of the solid and fluid phases.

$\rho_{RS}$, $\rho_{RF}$   material densities (real density) of the solid and fluid matrix.

$\rho_{RS0}$, $\rho_{RF0}$   initial densities of the solid and fluid matrix (material)

$\rho_F^L$, $\rho_F^c$   average densities of liquid solvent and solute in the fluid

$v_S$, $v_F$, $v$   velocities of the solid, fluid phase and the mixture average velocity.

$X_\alpha$   original position of $\alpha$-phase in the saturated porous media.

$X_F^\beta$   original position of $\beta$–constituent in the fluid.

$u_S$, $u_F$   diffusion velocities of the solid and fluid phases.

$v_F^\beta$   velocity of $\beta$–constituent in the fluid phase.

$u_F^L$, $u_F^c$   diffusion velocities of the liquid solvent and solute in the fluid phase.

$\hat{c}_S$, $\hat{c}_F$   mass exchange rate from other phase to solid phase and fluid phase.

$\hat{c}_F^L$, $\hat{c}_F^c$   mass exchange rate of the liquid solvent and solute in the fluid phase

$W_S$, $W_F$   diffusion velocities of the solid and fluid phase relative to solid phase.

$c_\mathrm{F}^\mathrm{L}$, $c_\mathrm{F}^\mathrm{c}$   mass fractions of the liquid solvent and solute in the fluid phase.

$\boldsymbol{\sigma}_\mathrm{S}$, $\boldsymbol{\sigma}_\mathrm{F}$,   Cauchy stress tensors of the solid and fluid phases.

$\boldsymbol{\sigma}$, $P_\mathrm{T}$   total Cauchy stress tensors and total pressure.

$P_\mathrm{S}$, $P_\mathrm{F}$   matrix (material) pressures of the solid and fluid ($P_\mathrm{F}$ is called pore pressure).

$\hat{\boldsymbol{p}}_\mathrm{S}$, $\hat{\boldsymbol{p}}_\mathrm{F}$   momentum supplies of the the solid and fluid phases.

$\boldsymbol{b}_\mathrm{S}$, $\boldsymbol{b}_\mathrm{F}$   external force densities of the the solid and fluid phases.

$\boldsymbol{a}_\mathrm{S}$, $\boldsymbol{a}_\mathrm{F}$   accelerations of the solid and fluid phases.

$\boldsymbol{q}_\alpha$, $r_\alpha$, $\hat{\varepsilon}_\alpha$   heat flux vector, external heat supply and energy supply of $\alpha$-phase, respectively.

$\xi_\mathrm{S}$, $\xi_\mathrm{F}$   internal energy densities of the solid and fluid phases.

$\tilde{\boldsymbol{\sigma}}$, $\tilde{\sigma}^\mathrm{p}$   Terzaghi's effective stress and dissipative effective pressure.

$\vartheta_\mathrm{S}$, $\vartheta_\mathrm{F}$   volumetric strain of the solid matrix (material) and fluid matrix (material).

$\varepsilon_c$, $\theta$   mass-exchange strain and temperature.

$\boldsymbol{D}_\mathrm{S}$, $\boldsymbol{D}_\mathrm{H}$   deformation rates of the solid phase and skeleton deformation rate

$\boldsymbol{F}_\mathrm{S}$, $\boldsymbol{F}_\mathrm{H}$   deformation gradients of the solid phase and solid skeleton.

$\boldsymbol{E}_\mathrm{S}$, $\boldsymbol{E}_\mathrm{H}$   Green strain of the solid phase and solid skeleton.

$\tilde{\boldsymbol{T}}_\mathrm{H}$   Kirchhoff-type Terzaghi effective stress of the solid skeleton.

$\eta_\mathrm{S}$, $\eta_\mathrm{F}$   entropy densities of the solid and fluid phases.

$\eta^r$, $\eta^i$   entropy flow and entropy production.

$\varsigma$   dissipation function.

$\boldsymbol{E}_\mathrm{H}^\mathrm{p}$, $\tilde{\boldsymbol{T}}_\mathrm{H}^\mathrm{p}$   Green plastic strain of the solid skeleton and Kirchhoff-type dissipative

Terzaghi effective stress.

$\bar{\mu}_F^L$, $\bar{\mu}_F^c$ chemical potentials with respects to $c_F^L$ and $c_F^c$ in the fluid phase.

$\mu_F^L$, $\mu_F^c$ chemical potentials of the liquid solvent and solute in the unit molar of fluid .

$\bar{\mu}_F^{c0}(\theta, P_0)$, $\bar{\mu}_F^{L0}(\theta, P_0)$ standard chemical potentials of NaCl solute and water solvent.

$M_F^L$, $M_F^c$ molar mass of the liquid solvent and solute in the fluid phase

$U_S$, $U_H$ displacement of the solid and the solid skeleton.

$\varepsilon_S$, $\varepsilon_S^p$ solid strain tensor and plastic strain tensor under infinitesimal strain condition.

$\varepsilon_{SV}$, $\varepsilon_{SV}^p$ solid volumetric strain and plastic volumetric strain of the solid.

$\varepsilon_{FV}$ volumetric strain of the fluid.

$\varepsilon_H$, $\varepsilon_H^p$ strain tensor and the plastic strain tensor of the solid skeleton under infinitesimal strain condition.

$\varepsilon_{HV}$, $\varepsilon_{HV}^p$ volumetric strain plastic volumetric strain of the solid skeleton.

$\psi_S$, $\psi_F$ Helmholtz free energy of the solid and fluid phases.

$\Psi_{HS}$, $\Psi_{HF}$ Helmholtz free energy due to the solid skeleton deformation and fluid absorb action.

$\Psi_{RS}$, $\Psi_{RF}$ Helmholtz free energy of the solid matrix and the fluid matrix.

$\varepsilon_{Hz}$, $\varepsilon_{Hz}^p$ vertical strain and the vertical plastic strain of the solid skeleton.

$\tilde{P}_z$, $\tilde{P}_z^p$ Terzaghi's vertical effective pressure and dissipative vertical effective pressure.

$P_{Sz}$, $\vartheta_{Sz}$ vertical pressure and strain of the solid matrix.

$\varepsilon_{Sz}$ vertical strain of the solid.

$E_{RS}$ compression modulus of the solid matrix.

$\tilde{P}_{z0}$, $\tilde{P}_{zref}$    initial vertical pressure and reference vertical pressure.

$\lambda^e$, $\lambda^p(c_F^c)$    rebound index and plastic compression index.

$\varepsilon_{Hmax}$    free swelling strain at $\tilde{P}_z = \tilde{P}_{z0}$ for de-ionized water.

$\omega$, $d_S$    water content and relative density of solid material.

$\varepsilon_{Hz}^e$    elastic vertical strain of the solid skeleton.

$\varepsilon_{Href}^p$    reference vertical plastic strain at $\tilde{P}_z = \tilde{P}_{zref}$

$\varepsilon_{Hz0}^p(c_F^c)$    initial vertical plastic strain at $\tilde{P}_z = \tilde{P}_{z0}$ and at the NaCl mass fraction $c_F^c$.

$\lambda_1^p$, $\lambda_2^p$, $\lambda_3^p$, $\lambda^c$    are three plastic parameters.

$\tilde{P}_{zc}(c_F^c)$,    yielding pressure at NaCl mass fraction $c_F^c$.

$\tilde{P}_{zc}(0)$    yielding pressure at $c_F^c = 0$ for de-ionized water.

$\varepsilon_{Hz}^{pc}(c_F^c)$    chemical plastic strain of the solid skeleton due to the change of mass fraction

$\overline{V}_{m,L}$    partial molar volume of liquid water.

$K_F$    volumetric modulus of the NaCl-water solution.

$\gamma_w$    gravity of water.

$k_{Fz}$, $k_d$    permeability coefficient and diffusion coefficient.

$\zeta_F$    Biot's increment of fluid content (fluid volume accumulation per unit bulk volume)

# 1. Introduction

Geotechnical Porous media include soil, rock, garbage, combustible ice, etc. When dealing with the geotechnical engineering involving environmental issues, the solid phase and pore fluid phase in the saturated porous media often contain some chemically active constituents (Di Maio et al., 2004; Chen et al., 2016; Detmann, 2021). The chemically active constituents will alter the microscopic molecular and ion structure of the porous media to a certain extent, and then exhibit the macroscopic changes in the physical, mechanical and hydraulic properties of the saturated porous media. These arise from molecule-to-molecule, molecule-to-ion, and ion-to-ion interactions are generally called chemical activity. The effects of chemical activity on the engineering mechanical and hydraulic properties of porous media can be relate to the fields such as nuclear industry, hazardous industry, municipal waste disposal, petroleum and gas extraction, and geo-environmental applications (Wei, 2014; Santamarina et al., 2018). For example, according to statistics, more than 290 landfills have produced excessive rubbish leachate in China since 2000, causing serious groundwater and soil pollution around the landfills and taking expensive costs of contaminated land restoration (Li, 2021). With the development of the environment geotechnical engineering, the influence of chemical activity on the mechanical and hydraulic behaviors of saturated porous media has attracted extensive attention. It has become a significant trend to study the role of chemical activity in geomechanics (Loret et al., 2002; Gajo et al., 2002; Hu, 2008; Thomas et al, 2010, 2012; Dominijanni et al., 2014; Guimarães et al., 2014; Chen et al., 2010; Ye et al., 2014; Xie et al., 2015; Santamarina et al., 2018; Bai et al., 2020).

When the solid matrix (called the solid particles in soil mechanics) is incompressible, the deformation of the solid phase in porous media is equal to the deformation caused by the change of porosity. In the paper, the deformation caused by the change of porosity is referred to as the solid skeleton deformation. Many scholars modified the expression of Terzaghi effective stress to study the interaction between chemical activity and solid skeleton deformation. For example, based on the electric charge conservation between the solid phase and the pore fluid phase, Dominijanni, (2014) improved the effective stress to establish theoretical models considering chemo-mechanical coupling. Based on the surface fractal characteristics of swelling porous media, Xu (2019) proposed a modified effective stress formula with osmotic suction to calculate the swelling deformation of bentonite in NaCl salt solution. Based on the surface potential energy, Wei (2014) and Ma et al. (2019) theoretically presented the formula of inter-granular stress to consider the chemo-mechanical coupling in the process of the establishment of constitutive model (Yang et al., 2022). Based on the experimental observations, Zhang et al (2020) proposed an effective stress equation with consideration of osmotic suction effects for the compacted GMZ01 bentonite specimen with a void ratio higher than 0.5. The above studies have effectively promoted the development of saturated geotechnical materials in the field of chemo-hydro-mechanics multi-field coupling, But the coupling of chemical activity and mechanic behavior under the compressible solid and fluid matrices remain to be further studied.

As an axiomatic theory, Mixture theory is often used to study the chemo-mechanical coupling effect of saturated porous media, especially when considering the compressibility of the solid matrix and the fluid matrix. Bowen (1976) established the classical mixture theory, in which the macroscopic form of the field equations including mass conservation, momentum conservation, and energy conservation is postulated. The hybrid mixture theory (abbreviated as HMT) introduced the averaging procedure into the mixture theory and established the relationship between the microscopic field equation and the macroscopic field equation (Hassanizadeh & Gray, 1979a, b; Bennethum, et al, 2000; Cushman et al., 2002). HMT (The hybrid mixture theory) found that the terms in Bowen's macroscopic field equation are precisely identified with microscopic counterparts after the microscopic field equations are averaged. To modified the concept of chemical potential defined by Bowen in the mixture theory and reveal the mechanical mechanism of swelling porous media, Bennethum (1996, 2000) defined a new chemical potential for a two-phase, $N$-constituent, swelling porous medium. Then she used free energy and the restrictions of the second thermodynamics law to establish a constitutive framework of swelling porous media. Borja (2006) derived the energy equation in terms of Skempton-type effective stress tensor and used the Skempton-type effective stress tensor to formulate the constitutive equation of saturated and partly saturated porous media. To highlight the effects of porosity on the constitutive characteristics of saturated porous media, Chen et al. (2016) and Ma et al. (2022) incorporated the chemical potential into Biot's theory of saturated porous media and established a thermal-hydro-mechanical-chemical coupling model in terms of total stress and pore pressure. Zhang (2017) and Zhang & Cheng, (2015) established a unified thermo-porosity-mechanical (TPM) coupled constitutive model of saturated clay based on non-equilibrium thermodynamics, and established a unified constitutive relationship using elastic potential energy and dissipation energy. The above theories have greatly promoted the development of mixture theory considering chemical activity from various aspects with breadth and depth.

Porosity reflects the basis structural characteristics of porous media and plays a key role in the mechanical properties, permeability and chemical properties of porous media. In the theory of porous media without considering the chemical activity, the porosity seriously affects the permeability and compressibility (drained bulk modulus) of saturated porous media, so the solid deformation is often divided into the deformations of the solid skeleton and matrix to establish the corresponding constitutive model (Carroll and Katsube, 1983; Detournay and Cheng, 1993; Cheng, 2016, Hu, 2016). Here, the volumetric deformation of the solid skeleton is related to the change in porosity, and the volumetric deformation of the solid matrix is defined as the volume change of the solid material. In the theory of porous media considering the chemical activity, the porosity seriously affects the mass fraction of each phase and the absorption effect of saturated porous media (Chapuis, 2004, Dehghani et al., 2018); it plays the same equally key role as the porous media without chemical activity. Therefore, in the theory of porous media considering chemical activity, it is equally necessary to separate the skeleton deformation from the solid deformations, and adopted the skeleton strain as an elementary variable to establish a chemical-mechanical coupling model of saturated porous media. However, to update, few theories divide the solid deformation into the solid skeleton and matrix deformations and use the solid skeleton and matrix strains as elementary variables to build the constitutive relationship of saturated porous media with chemical-activity.

Both in classical mixture theory (Bowen, 1976) and in the HMT ((Hassanizadeh & Gray, 1979a, b; Bennethum, et al, 2000; Cushman et al., 2002), the constitutive relation of saturated porous media includes those of the solid and fluid phases as well as those between the thermodynamic fluxes and forces. However, most of the current researches on the effect of chemical activity in the geotechnical field only focus on the constitutive model of the solid phase, without providing the constitutive model of the fluid phase and the constitutive models between the thermodynamic fluxes and force. In fact, in deducing the governing seepage equation of the fluid phase and the seepage-diffusion equation of solute, the constitutive model of the fluid phase and the constitutive models between the thermodynamic fluxes and force are inevitably used.

In addition, the different choices of chemical state variables lead to the different difficulties in establishing the fluid constitutive model and the entropy production ex-pression. The dissipation potentials arising from the entropy production expression seriously determine and restrict the constitutive models between the thermodynamic fluxes and force (De Groot & Mazur, 1962). Among mass fraction, mass concentration, chemical potential and osmotic pressure, it is found that the choice of the mass fraction as a chemical state variable is the easiest to formulate the second thermodynamic law and the entropy production. When the research object is a multiphase and multicomponent system, the mass fraction is conveniently used to derive the second law of thermodynamics and determine the value of chemical potential because the sum of all constituent mass fractions is equal to unity. Whereas the value of chemical potential determined by the mass concentration need to be constrained by the Gibbs-Duhem equation (De Groot & Mazur, 1962). The Gibbs-Duhem equation is complex in the multiphase and multicomponent system, especially in the case of porosity-dependent skeleton strain. The chemical potential changes with the mass fraction, the pressure and the temperature. The osmotic pressure occurs under the presence of a semipermeable membrane and the equilibrium of the second law of thermodynamics (Johnson et al., 2021). Since in the process of the consolidation of saturated porous media, the fluid phase cannot reach the thermodynamic equilibrium due to the seepage of fluid through the porous solid, in this case the thermodynamic mechanism of the osmotic pressure is implicitly contained in the seepage-diffusion equation and is difficult to determine it (can be explained in the following text). Considering the complex effecting factors between chemical potential and osmotic pres-sure, different from most of current constitutive models, the mass fraction is chosen as a chemical state variable to establish the constitutive model of saturated porous media with chemical activity in the paper.

The purpose of this paper is to establish a practical constitutive theoretical framework for considering the chemical activity of the geotechnical porous media based on the combination of HMT and the chemical thermodynamics. First, to highlight the important role of porosity in the chemo-hydro-mechanical coupling of porous media, the solid strain is decomposed into the solid matrix strain, the solid skeleton strain and the solid mass-exchange strain. The solid matrix strain is caused by the volumetric change of the solid material, and the solid skeleton volumetric strain is caused by the change in porosity, as well as the solid mass-exchange strain is caused by solid mass supply. Second, different from the previous researches, the matrix strain, the skeleton strain and the mass fraction of solute and their conjugate pairs (the effective stress, the matrix pressure and the chemical potential of solute) are adopted as the state variables of the constitutive model considering the coupling of chemical activity and porosity change. Third, the practical constitutive theoretical framework includes two parts: one is expressed by the free energy that is introduced to formulate the elastic model. The other is expressed by the dissipative potential that is introduced to formulate the irreversible constitutive model including the plastic model and the constitutive relationship between the thermodynamic fluxes and forces. Forth, under the guide of the above constitutive theoretical framework, a swelling constitutive model of bentonite consid-ering the effect of NaCl-solution chemical activity and the chemo-hydro-mechanical coupling is presented based on the HMT and the chemical thermodynamics. This model includes the constitutive equations of the solid and fluid phases as well as the constitutive equations between the thermodynamic fluxes and forces. The governing equation of consolidation for the solid and fluid phases and the seepage-diffusion equation of salt solute are derived from the above constitutive equations. The classical osmotic pressure can be deduced from the seepage-diffusion equation of solute and the thermodynamic equilibrium. This result shows that the classical osmotic pressure has been implied in the seepage-diffusion equation of solute. Finally, compared the theoretical simulation results with the existing experimental data, they are finely consistent with each other.

**2. The volume fraction and density of saturated porous media**

The saturated porous media consists of a solid phase and a fluid phase in the pores. For the practical purposes, the solid phase is simplified to contain only one constituent for ease of study. In order to exhibit physical and chemical activities, it is assumed that the fluid filling the pores is a solution, which contains a liquid solvent constituent and a solute constituent. In general, a mass exchange occurs between the solid and fluid phases and between the solvent and the solute due to physical and chemical activities. Let the symbol "S" be used to denote the solid phase and the symbol "F" denote the fluid phase; $\alpha=\{S, F\}$ be the phase indicating variable. Let $\varphi_\alpha$ be the volume fraction of $\alpha$-phase. $\rho_\alpha$ be the average density of $\alpha$-phase. $\rho_{R\alpha}$ be the mate-rial density of $\alpha$-phase, where $\rho_\alpha = \varphi_\alpha \rho_{R\alpha}$. The total density of the saturated porous medium mixture is $\rho = \rho_S + \rho_F$. It should be noted that the physical quantities and their field equations described on the macroscopic scale in the paper can be associated with those on the microscopic scale by mean of averaging procedure (Hassanizadeh and Gray, 1979a,b; Achanta, et al, 1997; Bennethum,1996; Bennethum, et al, 2000; Cushman et al., 2002). The combination of averaging method and mixture theory is referred to as hybrid mixture theory(HMT). By the definition of volume fraction, we have:

$$\varphi_S + \varphi_F = 1 \tag{1}$$

As mentioned above, the fluid phase contains two constituents: liquid solvent and solute. Because the liquid solvent and the solute are mixed together, it is difficult to isolate their respective volumes and volume fractions from the fluid phase. Thus, the liquid solvent and the solute have the same volume as the fluid phase. Let the symbol "L" denote the liquid solvent and the symbol "c" denote the solute; $\beta=\{L, c\}$ be the constituent indicating variable in the fluid phase; $\rho_F^\beta$ be the average density of $\beta$–constituent in the fluid. By the definition of average density, we have:

$$\rho_F = \sum_{\beta=L,c} \rho_F^\beta = \rho_F^L + \rho_F^c \tag{2}$$

## 3. Conservation equations

### 3.1. Conservation of mass

The hybrid mixture theory averages the solid phase and the fluid phase over the representative element volume (REV) of the mixture so that the solid phase and the fluid phase continuously share the space position of the mixture. If the original position of α-phase in the saturated porous media is denoted by $X_\alpha$, and its spatial position at time $t$ by $x$, the motion equation of each phase can be expressed as:

$$x = x_\alpha(X_\alpha, t) \tag{3}$$

If $v_\alpha$ is the velocity of α-phase, the phase-averaged velocity of the saturated porous media is:

$$v = \frac{1}{\rho}(\rho_S v_S + \rho_F v_F) \tag{4}$$

The relative velocity of α-phase is defined as:

$$u_\alpha = v_\alpha - v \tag{5}$$

For scalar or vector fields $\Gamma$ defined on $x$ and $t$, the material derivative with respect to the α-phase is defined as:

$$\frac{d^\alpha \Gamma}{dt} = \frac{\partial \Gamma}{\partial t} + \nabla \Gamma \cdot v_\alpha \tag{6}$$

The material derivative of $\Gamma$ with respect to the whole mixture is defined as:

$$\frac{d\Gamma}{dt} = \frac{\partial \Gamma}{\partial t} + \nabla \Gamma \cdot v \tag{7}$$

Similar to the α-phase, if the original position of the β-constituent in the fluid is denoted by $X_F^\beta$, and its spatial position at time $t$ by $x$, the motion equation of the β-constituent in the fluid phase can be expressed as:

$$x = x_F^\beta(X_F^\beta, t) \tag{8}$$

If $v_F^\beta$ is the velocity of the β-constituent in the fluid phase, for scalar or vector fields $\Gamma$ defined on $x$ and $t$, the material derivative with respect to the β-constituent in the

fluid is defined as:

$$\frac{d_F^\beta \boldsymbol{\Gamma}}{dt} = \frac{\partial \boldsymbol{\Gamma}}{\partial t} + \nabla \boldsymbol{\Gamma} \cdot \boldsymbol{v}_F^\beta \tag{9}$$

The relationship between $\boldsymbol{v}_F$ and the velocity $\boldsymbol{v}_F^\beta$ of the $\beta$-constituent in the fluid is

$$\boldsymbol{v}_F = \frac{1}{\rho_F} \left( \rho_F^L \boldsymbol{v}_F^L + \rho_F^c \boldsymbol{v}_F^c \right) \tag{10}$$

The relative velocity of the $\beta$-constituent with respects to the fluid phase is defined as:

$$\boldsymbol{u}_F^\beta = \boldsymbol{v}_F^\beta - \boldsymbol{v}_F \tag{11}$$

Since we are assuming that the mass exchange rate from other phase to $\alpha$-phase is $\hat{c}_\alpha$, the conservation of mass for the solid and fluid phases are (Bowen, 1976; Hassanizadeh and Gray, 1979a,b; Bennethum et al, 2000):

$$\left. \begin{aligned} \frac{d^S \rho_S}{dt} + \rho_S \nabla \cdot \boldsymbol{v}_S &= \hat{c}_S \\ \frac{d^F \rho_F}{dt} + \rho_F \nabla \cdot \boldsymbol{v}_F &= \hat{c}_F \end{aligned} \right\} \tag{12}$$

where $\hat{c}_S + \hat{c}_F = 0$ (Bowen, 1976). Combining Eqs. (4) and (12) yields:

$$\frac{d\rho}{dt} + \rho \nabla \cdot \boldsymbol{v} = 0 \tag{13}$$

In geotechnical mechanics, engineers are concerned with the deformation of the solid phase rather than the deformation of the entire saturated porous media mixture, so the solid deformation is used as the reference configuration for poromechanics when building constitutive model (Houlsby. 1997; Coussy, 2004; Borja, 2006). Let the relative velocity of the fluid phase to the solid phase be defined as $\boldsymbol{W}_F = \boldsymbol{v}_F - \boldsymbol{v}_S$. Substituting the definition of $\boldsymbol{W}_F$, $\rho_\alpha = \varphi_\alpha \rho_{R\alpha}$, and Eq. (4) into Eq.(12) yields (Borja, 2006):

$$\left.\begin{array}{l}\dfrac{\varphi_{S}}{\rho_{RS}}\dfrac{d^{S}\rho_{RS}}{dt}+\dfrac{d^{S}\varphi_{S}}{dt}+\varphi_{S}\nabla\cdot\boldsymbol{v}_{S}-\dfrac{\hat{c}_{S}}{\rho_{RS}}=0\\ \dfrac{\varphi_{F}}{\rho_{RF}}\dfrac{d^{F}\rho_{RF}}{dt}+\dfrac{d^{S}\varphi_{F}}{dt}+\varphi_{F}\nabla\cdot\boldsymbol{v}_{S}+\varphi_{F}\nabla\cdot\boldsymbol{W}_{F}+\boldsymbol{W}_{F}\cdot\nabla\varphi_{F}-\dfrac{\hat{c}_{F}}{\rho_{RF}}=0\end{array}\right\} \quad (14)$$

If the mass exchange rate from other constituent to the $\beta$-constituent of the fluid phase is denoted by $\hat{c}_F^\beta$, the mass conservations of the $\beta$-constituent in the fluid phase are (Hassanizadeh and Gray, 1979a,b; Bennethum et al, 2000):

$$\left.\begin{array}{l}\dfrac{d_F^L \rho_F^L}{dt}+\rho_F^L \nabla\cdot\boldsymbol{v}_F^L=\hat{c}_F^L\\ \dfrac{d_F^c \rho_F^c}{dt}+\rho_F^c \nabla\cdot\boldsymbol{v}_F^c=\hat{c}_F^c\end{array}\right\} \quad (15)$$

Obviously, Eq. $(12)_2$ can be obtained from Eqs. (10) and (15), so $\hat{c}_F^L+\hat{c}_F^c=\hat{c}_F$. Let $c_F^\beta=\rho_F^\beta/\rho_F$ is the mass fraction of the $\beta$-constituent in the fluid phase, using Eqs.$(12)_2$ and (15) yields (Bennethum et al, 2000):

$$\left.\begin{array}{l}\rho_F \dfrac{d_F^L c_F^L}{dt}+\nabla\cdot(\rho_F^L \boldsymbol{u}_F^L)=\hat{c}_F^L-c_F^L \hat{c}_F\\ \rho_F \dfrac{d_F^c c_F^c}{dt}+\nabla\cdot(\rho_F^c \boldsymbol{u}_F^c)=\hat{c}_F^c-c_F^c \hat{c}_F\end{array}\right\} \quad (16)$$

*3.2. Conservation of momentum and moment of momentum*

Let $\boldsymbol{\sigma}_\alpha$ ($\alpha\in\{S,F\}$) be the Cauchy stress tensor of $\alpha$-phase in the saturated porous media. Cauchy stress tensor is positive in tension in the paper. Because the deformation and destroy are only related to the solid phase in the geotechnical engineers, it is only necessary to build the constitutive model of the constituent rather than the entire saturated porous media mixture. Therefore, there is no need for the total stress tensor of the entire mixture as defined by the classical mixture theory. Only the total Cauchy stress tensor $\boldsymbol{\sigma}$ of the solid phase is introduced, it is defined

as (equal to the main part of the total stress tensor of the entire mixture in the classical mixture):

$$\boldsymbol{\sigma} = \boldsymbol{\sigma}_S + \boldsymbol{\sigma}_F \tag{17}$$

It is noted that pressure is positive in compression in geotechnical mechanics. The total pressure is defined as $P_T = -\boldsymbol{\sigma} : \boldsymbol{I} / 3$. The pressure $P_S$ of the solid matrix (material) is defined as $P_S = -\boldsymbol{\sigma}_S : \boldsymbol{I} / (3\varphi_S)$. Assuming that water viscosity is not taken into account, the relationship between the pressure $P_F$ of the fluid matrix (material) and $\boldsymbol{\sigma}_F$ is $P_F = -\boldsymbol{\sigma}_F : \boldsymbol{I} / (3\varphi_F)$. $P_F$ is also called pore pressure. Using the above relationship, Eq. (17) gives:

$$P_T = \varphi_S P_S + \varphi_F P_F \tag{18}$$

Fig.1 illustrates the relationship of various pressures acting on the representative elementary volume expressed in Eq. (18).

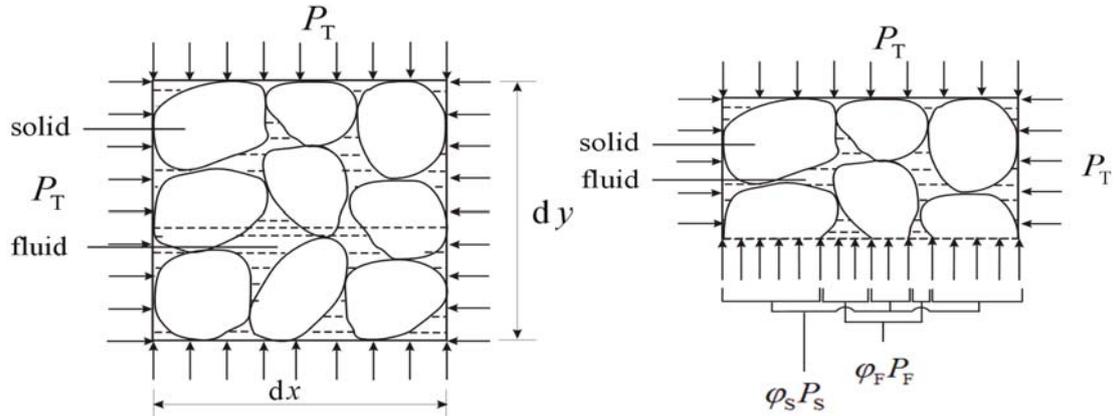

**Fig. 1** The schematic diagram for the representative elements body of saturated porous media

According to the mixture theory, the momentum conservation equations of two phases are (Hassanizadeh and Gray, 1979a,b; Bennethum et al, 2000):

$$\left. \begin{array}{l} \rho_S \boldsymbol{a}_S = \nabla \cdot \boldsymbol{\sigma}_S + \rho_S \boldsymbol{b}_S + \hat{\boldsymbol{p}}_S \\ \rho_F \boldsymbol{a}_F = \nabla \cdot \boldsymbol{\sigma}_F + \rho_F \boldsymbol{b}_F + \hat{\boldsymbol{p}}_F \end{array} \right\} \tag{19}$$

where $\hat{\boldsymbol{p}}_\alpha$ is the momentum supply of α-phase. $\boldsymbol{b}_\alpha$ is the external force density of α-phase, $\boldsymbol{a}_\alpha$ is the acceleration of α-phase. According to HMT (Bowen, 1976; Hassanizadeh and Gray, 1979a,b; Bennethum et al, 2000), we have:

$$\sum_{S,F}(\hat{\boldsymbol{p}}_\alpha + \hat{c}_\alpha \boldsymbol{u}_\alpha) = 0 \tag{20}$$

Assuming that the momentum moment supply vector of α-phase is zero, the stress tensor $\boldsymbol{\sigma}_\alpha$ is symmetric according to the momentum moment equilibrium of α-phase.

Although the fluid phase contains two constituents: liquid solvents and solutes, since the solvents and the solutes are mixed together, it is difficult to distinguish their respective stress tensor in the laboratory and the field. Therefore, as far as reality is concerned, we only provide the conversation equations of momentum and momentum moment of the solid and fluid phases.

*3.3. The energy balance equation*

Let $\boldsymbol{q}_\alpha, r_\alpha$ and $\hat{\varepsilon}_\alpha$ be the heat flux vector, external heat supply and energy supply of α-phase, respectively. Denoting the internal energy density of α-phase by $\xi_\alpha$, then the internal energy equations of both the solid and fluid phases can be expressed as (Bowen, 1976; Hassanizadeh and Gray, 1979a,b; Bennethum et al, 2000)

$$\left.\begin{aligned}\rho_S \frac{d^S \xi_S}{dt} &= \boldsymbol{\sigma}_S : \nabla \boldsymbol{v}_S - \nabla \cdot \boldsymbol{q}_S + \rho_S r_S + \hat{\varepsilon}_S \\ \rho_F \frac{d^F \xi_F}{dt} &= \boldsymbol{\sigma}_F : \nabla \boldsymbol{v}_F - \nabla \cdot \boldsymbol{q}_F + \rho_F r_F + \hat{\varepsilon}_F + \sum_{\beta=L,c} \rho_F^\beta \boldsymbol{u}_F^\beta \cdot \boldsymbol{b}_F^\beta \end{aligned}\right\} \tag{21}$$

According to the classical mixture theory (Bowen, 1976; Hassanizadeh and Gray, 1979a,b; Bennethum et al, 2000), $\hat{\varepsilon}_\alpha$ in Eq. (21) satisfies:

$$\sum_{\alpha=S,F}\left[\hat{\varepsilon}_\alpha + \boldsymbol{u}_\alpha \cdot \hat{\boldsymbol{p}}_\alpha + \hat{c}_\alpha(\xi_\alpha + \frac{1}{2}u_\alpha^2)\right]=0 \qquad (22)$$

In the porous media, the porosity represents the microscopic structure. For example, the free swelling capacity of swelling soils depends on the water content, which is closely related to the porosity under saturated conditions (Bennethum, et al, 2000; Cushman et al., 2002). Therefore, it is necessary to isolate the porosity-dependent strain from the solid strain. The porosity-dependent strain is called the skeleton strain in the geotechnical mechanics. However, $\nabla \boldsymbol{v}_S$ depend on not only the skeleton strain but also the solid matrix strain, there is no one-to-one relationship between $\nabla \boldsymbol{v}_S$ in Eq. (21) and the skeleton strain. To analyze the key role of porosity on constitutive model, the solid and fluid strain need to be divided into the skeleton strain and the matrix strain using Eqs. (14), (17), (21) and (22).

More specific, there is $\boldsymbol{\sigma}_F = -\varphi_F P_F \boldsymbol{I}$ when neglecting the fluid viscosity. Combining this equation, the definition of $\boldsymbol{W}_F$ and Eq. (17), the sum of the strain energy rates of all phases in Eq.(21) can be obtained:

$$\sum_{\alpha=S,F} \boldsymbol{\sigma}_\alpha : \nabla \boldsymbol{v}_\alpha = \boldsymbol{\sigma} : \nabla \boldsymbol{v}_\alpha - \varphi_F P_F \nabla \cdot \boldsymbol{W}_F \qquad (23)$$

Substituting Eq. (14)$_2$ into Eq. (23) gives

$$\begin{aligned}\sum_{\alpha=S,F} \boldsymbol{\sigma}_\alpha : \nabla \boldsymbol{v}_\alpha &= \boldsymbol{\sigma} : \nabla \boldsymbol{v}_S + P_F \frac{d^S \varphi_F}{dt} + P_F \varphi_F \nabla \cdot \boldsymbol{v}_S \\ &+ P_F \boldsymbol{W}_F \cdot \nabla \varphi_F + P_F \frac{\varphi_F}{\rho_{RF}} \frac{d^F \rho_{RF}}{dt} - P_F \frac{\hat{c}_F}{\rho_{RF}}\end{aligned} \qquad (24)$$

Let $\tilde{\boldsymbol{\sigma}} = \boldsymbol{\sigma} + P_F \boldsymbol{I}$, which expression is called the Terzaghi's effective stress in the geotechnical mechanics. Substituting Eqs. (1) and (14)$_1$ into Eq.(24) and replacing

$\boldsymbol{\sigma} + P_F \boldsymbol{I}$ by $\tilde{\boldsymbol{\sigma}}$ to simply it, Eq. (24) yields:

$$\sum_{\alpha=S,F} \boldsymbol{\sigma}_\alpha : \nabla \boldsymbol{v}_\alpha = \tilde{\boldsymbol{\sigma}} : \nabla \boldsymbol{v}_S + P_F \frac{\varphi_S}{\rho_{RS}} \frac{d^S \rho_{RS}}{dt} \\ + P_F \boldsymbol{W}_F \cdot \nabla \varphi_F + P_F \frac{\varphi_F}{\rho_{RF}} \frac{d^F \rho_{RF}}{dt} - P_F (\frac{\hat{c}_S}{\rho_{RS}} + \frac{\hat{c}_F}{\rho_{RF}}) \quad (25)$$

The matrix volumetric strain of $\alpha$-phase is defined as $\vartheta_\alpha = \ln(\rho_{R\alpha}/\rho_{R\alpha 0})$, where $\rho_{R\alpha 0}$ is the initial density of $\alpha$-phase. $\vartheta_\alpha$ is positive in compression. According to the definition of $\vartheta_\alpha$ and $\alpha = F$, the second item $\varphi_S P_F d^S \rho_{RS}/(\rho_{RS} dt)$ on the right side of Eq. (25) becomes $\varphi_S P_F d^S \vartheta_S / dt$, which shows that the volumetric strain of the solid matrix is energy-conjugated with the pore pressure of the fluid phase. However, it is well known that the strain of the solid matrix should depend on the pressure of the solid matrix (Geertma, 1957, Cheng, 2016). This indicates that if the pore pressure is selected to energy-conjugate with the strain of the solid matrix in Eq.(25), it does not well capture the mesoscopic deformation characteristics of the solid matrix, and also not conducive to the establishment of a constitutive model of saturated porous media. That is, the second item on the right side of Eq. (25) should change the energy-conjugated pair. Now we carry out it.

Based on Eq.(18), the pressure $P_S$ acting on the solid matrix can be expressed as:

$$P_S = \frac{P_T - \varphi_F P_F}{\varphi_S} \quad (26)$$

Using Eq.(26) and $\vartheta_\alpha = \ln(\rho_{R\alpha}/\rho_{R\alpha 0})$, Eq. (25) become:

$$\sum_{\alpha=S,F} \boldsymbol{\sigma}_\alpha : \nabla \boldsymbol{v}_\alpha = \tilde{\boldsymbol{\sigma}} : (\nabla \boldsymbol{v}_S + \frac{d^S \vartheta_S}{dt} \frac{\boldsymbol{I}}{3}) + \sum_\alpha \varphi_\alpha P_\alpha \frac{d^S \vartheta_\alpha}{dt}$$
$$+ P_F \boldsymbol{W}_F \cdot \nabla \varphi_F - P_F (\frac{\hat{c}_S}{\rho_{RS}} + \frac{\hat{c}_F}{\rho_{RF}}) \tag{27}$$

However, from (14)$_1$, $\nabla \boldsymbol{v}_S + d^S \vartheta_S \boldsymbol{I}/(3dt)$ is related to not only the porosity but also the exchange of mass resulted from physical migration or chemical action. If the deformation rate resulted from mass exchange is abbreviated as the mass-exchange deformation rate and denoted by $d\varepsilon_c/dt$, which is defined as:

$$\frac{d\varepsilon_c}{dt} = \frac{\hat{c}_S}{\rho_S} \tag{28}$$

where $\varepsilon_c$ is the mass-exchange strain. Since we require the skeleton volumetric strain uniquely determined by the change in porosity, based on Eq.(14)$_1$, the term $\nabla \boldsymbol{v}_S + d^S \vartheta_S \boldsymbol{I}/(3dt)$ still need minus the mass-exchange deformation rate in order to become the skeleton deformation rate. Let the symbol $\boldsymbol{D}_S$ denote the deformation rate of the solid phase, and $\boldsymbol{D}_H$ denote the deformation rate of the solid skeleton. $\boldsymbol{D}_S$ is defined as

$$\boldsymbol{D}_S = \frac{\left[\nabla \boldsymbol{v}_S + (\nabla \boldsymbol{v}_S)^T\right]}{2} \tag{29}$$

Then, $\boldsymbol{D}_H$ is defined as

$$\boldsymbol{D}_H = \boldsymbol{D}_S + \frac{1}{3} \frac{d^S \vartheta_S}{dt} \boldsymbol{I} - \frac{1}{3} \frac{d\varepsilon_c}{dt} \boldsymbol{I} \tag{30}$$

According to Eqs. (29) and (30) and using Eq. (14)$_1$, there are

$$\boldsymbol{D}_S - \frac{1}{3} \operatorname{tr} \boldsymbol{D}_S \boldsymbol{I} = \boldsymbol{D}_H - \frac{1}{3} \operatorname{tr} \boldsymbol{D}_H \boldsymbol{I} \tag{31}$$

$$\operatorname{tr} \boldsymbol{D}_{\mathrm{H}} = \operatorname{tr} \boldsymbol{D}_{\mathrm{S}} + \frac{\mathrm{d}^{\mathrm{S}} \vartheta_{\mathrm{S}}}{\mathrm{d}t} - \frac{\mathrm{d}\varepsilon_{\mathrm{c}}}{\mathrm{d}t} = -\frac{1}{\varphi_{\mathrm{S}}} \frac{\mathrm{d}^{\mathrm{S}} \varphi_{\mathrm{S}}}{\mathrm{d}t} \tag{32}$$

Eq. (32) shows that $\operatorname{tr} \boldsymbol{D}_{\mathrm{H}}$ is uniquely determined by the solid volume fraction that has a one to one relationship with porosity. In geotechnical mechanics, a strain related to the change in the porosity is called the skeleton strain, that is why $\boldsymbol{D}_{\mathrm{H}}$ is called the skeleton deformation rate in the paper. Since $\operatorname{tr} \boldsymbol{D}_{\mathrm{H}}$ depends uniquely on the change in the porosity, it is relatively easier to establish the constitutive model using $\boldsymbol{D}_{\mathrm{H}}$ than using $\nabla \boldsymbol{v}_{\mathrm{S}} + \mathrm{d}^{\mathrm{S}} \vartheta_{\mathrm{S}} \boldsymbol{I} / (3 \mathrm{d}t)$ or $\boldsymbol{D}_{\mathrm{S}}$. After replacing $\nabla \boldsymbol{v}_{\mathrm{S}} + \mathrm{d}^{\mathrm{S}} \vartheta_{\mathrm{S}} \boldsymbol{I} / (3 \mathrm{d}t)$ by $\boldsymbol{D}_{\mathrm{H}}$ and using the definition expression of $P_{\mathrm{S}}$, Eq.(27) becomes:

$$\sum_{\alpha=\mathrm{S,F}} \boldsymbol{\sigma}_\alpha : \nabla \boldsymbol{v}_\alpha = \tilde{\boldsymbol{\sigma}} : \boldsymbol{D}_{\mathrm{H}} + \sum_{\alpha=\mathrm{S,F}} \varphi_\alpha P_\alpha \frac{\mathrm{d}^{\mathrm{S}} \vartheta_\alpha}{\mathrm{d}t} + P_{\mathrm{F}} \boldsymbol{W}_{\mathrm{F}} \cdot \nabla \varphi_{\mathrm{F}} - (\frac{P_{\mathrm{S}}}{\rho_{\mathrm{RS}}} \hat{c}_{\mathrm{S}} + \frac{P_{\mathrm{F}}}{\rho_{\mathrm{RF}}} \hat{c}_{\mathrm{F}}) \tag{33}$$

Let $\boldsymbol{W}_\alpha = \boldsymbol{u}_\alpha - \boldsymbol{u}_{\mathrm{S}}$, adding all of Eq. (21), then substituting Eqs. (22) and (33) into it yields:

$$\sum_{\alpha=\mathrm{S,F}} \rho_\alpha \frac{\mathrm{d}^\alpha \xi_\alpha}{\mathrm{d}t} = \tilde{\boldsymbol{\sigma}} : \boldsymbol{D}_{\mathrm{H}} + \sum_{\alpha=\mathrm{S,F}} \varphi_\alpha P_\alpha \frac{\mathrm{d}^{\mathrm{S}} \vartheta_\alpha}{\mathrm{d}t} + \sum_{\alpha=\mathrm{S,F}} (-\nabla \cdot \boldsymbol{q}_\alpha + \rho_\alpha r_\alpha)$$

$$+ \boldsymbol{W}_{\mathrm{F}} \cdot (P_{\mathrm{F}} \nabla \varphi_{\mathrm{F}} - \hat{\boldsymbol{p}}_{\mathrm{F}}) - \sum_{\alpha=\mathrm{S,F}} \hat{c}_\alpha \left( \xi_\alpha + \frac{1}{2} \boldsymbol{W}_\alpha^2 + \frac{P_\alpha}{\rho_{\mathrm{R}\alpha}} \right) + \sum_{\beta=\mathrm{L,c}} \rho_{\mathrm{F}}^\beta \boldsymbol{u}_{\mathrm{F}}^\beta \cdot \boldsymbol{b}_{\mathrm{F}}^\beta \tag{34}$$

Let $\boldsymbol{F}_{\mathrm{S}} = \partial \boldsymbol{x} / \partial \boldsymbol{X}_{\mathrm{S}}$, $\boldsymbol{E}_{\mathrm{S}} = \frac{1}{2} (\boldsymbol{F}_{\mathrm{S}}^{\mathrm{T}} \cdot \boldsymbol{F}_{\mathrm{S}} - \boldsymbol{I})$, $J_{\mathrm{S}} = \det[\partial \boldsymbol{x} / \partial \boldsymbol{X}_{\mathrm{S}}]$, $J_{\mathrm{RS}} = \rho_{\mathrm{RS}} / \rho_{\mathrm{RS0}}$, $J_{\mathrm{c}} = \exp(\int_0^t \hat{c}_{\mathrm{S}} / \rho_{\mathrm{S}} \mathrm{d}t)$, $\boldsymbol{F}_{\mathrm{H}} = [J_{\mathrm{c}} / J_{\mathrm{RS}}]^{1/3} \boldsymbol{F}_{\mathrm{S}}$, $\boldsymbol{E}_{\mathrm{H}} = \frac{1}{2} (\boldsymbol{F}_{\mathrm{H}}^{\mathrm{T}} \cdot \boldsymbol{F}_{\mathrm{H}} - \boldsymbol{I})$, $\tilde{\boldsymbol{T}}_{\mathrm{H}} = \boldsymbol{F}_{\mathrm{H}}^{-1} \cdot \tilde{\boldsymbol{\sigma}} \cdot \boldsymbol{F}_{\mathrm{H}}^{-\mathrm{T}}$. Where $\boldsymbol{F}_{\mathrm{S}}$ is the deformation gradient of the solid phase, $\boldsymbol{E}_{\mathrm{S}}$ is the Green strain of the solid phase, $\boldsymbol{F}_{\mathrm{H}}$ is the deformation gradient of the solid skeleton, $\boldsymbol{E}_{\mathrm{H}}$ is the Green strain of the solid skeleton, $\tilde{\boldsymbol{T}}_{\mathrm{H}}$ is the Kirchhoff-type Terzaghi effective stress of the solid skeleton. Substituting $\tilde{\boldsymbol{T}}_{\mathrm{H}}$ and $\boldsymbol{E}_{\mathrm{H}}$ into Eq. (34) gives

$$\sum_{\alpha=\mathrm{S},\mathrm{F}} \rho_\alpha \frac{\mathrm{d}^\alpha \xi_\alpha}{\mathrm{d}t} = (\tilde{\boldsymbol{T}}_{\mathrm{H}} : \frac{\mathrm{d}^{\mathrm{S}} \boldsymbol{E}_{\mathrm{H}}}{\mathrm{d}t} + \varphi_{\mathrm{S}} P_{\mathrm{S}} \frac{\mathrm{d}^{\mathrm{S}} \vartheta_{\mathrm{S}}}{\mathrm{d}t}) + \varphi_{\mathrm{F}} P_{\mathrm{F}} \frac{\mathrm{d}^{\mathrm{F}} \vartheta_{\mathrm{F}}}{\mathrm{d}t} + \sum_{\alpha=\mathrm{S},\mathrm{F}} (-\nabla \cdot \boldsymbol{q}_\alpha + \rho_\alpha r_\alpha)$$
$$+ \boldsymbol{W}_{\mathrm{F}} \cdot (P_{\mathrm{F}} \nabla \varphi_{\mathrm{F}} - \hat{\boldsymbol{p}}_{\mathrm{F}}) - \sum_{\alpha=\mathrm{S},\mathrm{F}} \hat{c}_\alpha \left( \frac{P_\alpha}{\rho_{\mathrm{R}\alpha}} + \frac{1}{2} W_\alpha^2 + \xi_\alpha \right) + \sum_{\beta=\mathrm{L},\mathrm{c}} \rho_{\mathrm{F}}^\beta \boldsymbol{u}_{\mathrm{F}}^\beta \cdot \boldsymbol{b}_{\mathrm{F}}^\beta \quad (35)$$

Eq. (34) can be also rewritten as follows:

$$\sum_{\alpha=\mathrm{S},\mathrm{F}} \rho_\alpha \frac{\mathrm{d}^\alpha \xi_\alpha}{\mathrm{d}t} = \rho_{\mathrm{S}} (\frac{\tilde{\boldsymbol{\sigma}}}{\rho_{\mathrm{S}}} : \boldsymbol{D}_{\mathrm{H}} + \frac{P_{\mathrm{S}}}{\rho_{\mathrm{RS}}} \frac{\mathrm{d}^{\mathrm{S}} \vartheta_{\mathrm{S}}}{\mathrm{d}t}) + \rho_{\mathrm{F}} \frac{P_{\mathrm{F}}}{\rho_{\mathrm{RF}}} \frac{\mathrm{d}^{\mathrm{F}} \vartheta_{\mathrm{F}}}{\mathrm{d}t} + \sum_{\alpha=\mathrm{S},\mathrm{F}} (-\nabla \cdot \boldsymbol{q}_\alpha + \rho_\alpha r_\alpha)$$
$$+ \boldsymbol{W}_{\mathrm{F}} \cdot (P_{\mathrm{F}} \nabla \varphi_{\mathrm{F}} - \hat{\boldsymbol{p}}_{\mathrm{F}}) - \sum_{\alpha=\mathrm{S},\mathrm{F}} \hat{c}_\alpha \left( \frac{P_\alpha}{\rho_{\mathrm{R}\alpha}} + \frac{1}{2} W_\alpha^2 + \xi_\alpha \right) + \sum_{\beta=\mathrm{L},\mathrm{c}} \rho_{\mathrm{F}}^\beta \boldsymbol{u}_{\mathrm{F}}^\beta \cdot \boldsymbol{b}_{\mathrm{F}}^\beta \quad (36)$$

## 4. Entropy flow and production

From Eq. (30), it can be seen that the solid deformation rate $\boldsymbol{D}_{\mathrm{S}}$ of the solid phase is divided into three parts: the volumetric deformation rate of solid matrix $\mathrm{d}^{\mathrm{S}} \vartheta_{\mathrm{S}} / \mathrm{d}t$, the skeleton deformation rate $\boldsymbol{D}_{\mathrm{H}}$ and the mass-exchange deformation rate $\mathrm{d} \varepsilon_{\mathrm{c}} / \mathrm{d}t$. From Eq. (35), it can be seen that $\mathrm{d}^{\mathrm{S}} \boldsymbol{E}_{\mathrm{H}} / \mathrm{d}t$ and $\mathrm{d}^{\mathrm{S}} \vartheta_{\mathrm{S}} / \mathrm{d}t$ as well as the volumetric deformation rate of the fluid matrix $\mathrm{d}^{\mathrm{F}} \vartheta_{\mathrm{F}} / \mathrm{d}t$ are energy-conjugated with $\tilde{\boldsymbol{T}}_{\mathrm{H}}$, $\varphi_{\mathrm{S}} P_{\mathrm{S}}$ and $\varphi_{\mathrm{F}} P_{\mathrm{F}}$, respectively. According to the thermodynamic theory (Annamalai et al., 2011), in general, the Green strain $\boldsymbol{E}_{\mathrm{H}}$ of the solid skeleton, the volumetric strain $\vartheta_{\mathrm{S}}$ of the solid matrix and the volumetric strain $\vartheta_{\mathrm{F}}$ of the fluid matrix should be selected as the strain state variables of the constitutive model. The Kirchhoff-type Terzaghi effective stress $\tilde{\boldsymbol{T}}_{\mathrm{H}}$, the pressure $P_{\mathrm{S}}$ of the solid matrix and the pressure $P_{\mathrm{F}}$ of the fluid matrix (i.e. pore pressure) are selected as the stress state variables of the constitutive model. In addition, according to thermodynamics (Annamalai et al., 2011), the mass fractions of constituents and their energy-conjugated pairs (chemical potentials) in each phase are still selected as the state variables of the constitutive model when the chemical activity is taken into accounts. Since there is heterogeneity between the solid and the fluid phases in saturated porous media, the internal energy of $\alpha$-phase is simplified to only depend on the mass fraction of the constituents within the same phase (Bennethum et al, 2000). For geomaterials, plasticity is a frequent mechanical phenomenon. When the constitutive mode need to describe a plastic behavior, an internal strain variable (generally called the plastic strain) and its energy-conjugated pair (generally called the dissipative stress) are introduced in the expression of internal energy (Collins and Houlsby,1997). According to the above explanation and assuming that a common temperature $\theta$ is assigned for both phase, i.e. $\theta = \theta_S = \theta_F$, the internal energies of the solid and fluid phases in Eq. (35) can be expressing as $\xi_S(\eta_S, \boldsymbol{E}_H, \boldsymbol{E}_H^p, \vartheta_S)$ and $\xi_F(\eta_F, \boldsymbol{E}_H, \boldsymbol{E}_H^p, \vartheta_F, c_F^L, c_F^c)$. Where, $\eta_\alpha$ is the entropy density of $\alpha$-phase, $\boldsymbol{E}_H^p$ is the Green plastic strain of the solid skeleton, $c_F^L = \rho_F^L / \rho_F$ and $c_F^c = \rho_F^c / \rho_F$ are the mass fraction of liquid solvent and solute, respectively. Completely differentiating the internal energies of the solid and fluid phases yields:

$$\sum_{\alpha=S,F} \rho_\alpha \frac{d^\alpha \xi_\alpha}{dt} = \rho_S \frac{\partial \xi_S}{\partial \eta_S} \frac{d^S \eta_S}{dt} + \rho_F \frac{\partial \xi_F}{\partial \eta_F} \frac{d^F \eta_F}{dt} + (\rho_S \frac{\partial \xi_S}{\partial \boldsymbol{E}_H} + \rho_F \frac{\partial \xi_F}{\partial \boldsymbol{E}_H}) : \frac{d^S \boldsymbol{E}_H}{dt} +$$
$$(\rho_S \frac{\partial \xi_S}{\partial \boldsymbol{E}_H^p} + \rho_F \frac{\partial \xi_F}{\partial \boldsymbol{E}_H^p}) \frac{d^S \boldsymbol{E}_H^p}{dt} + \rho_S \frac{\partial \xi_S}{\partial \vartheta_S} \frac{d^S \vartheta_S}{dt} + \rho_F \frac{\partial \xi_F}{\partial \vartheta_F} \frac{d^F \vartheta_F}{dt} + \rho_F \frac{\partial \xi_F}{\partial c_F^L} \frac{d^F c_F^L}{dt} + \quad (37)$$
$$\rho_F \frac{\partial \xi_F}{\partial c_F^c} \frac{d^F c_F^c}{dt} - (\rho_F \frac{\partial \xi_F}{\partial \boldsymbol{E}_H} : \nabla \boldsymbol{E}_H + \rho_F \frac{\partial \xi_F}{\partial \boldsymbol{E}_H^p} : \nabla \boldsymbol{E}_H^p) \cdot \boldsymbol{W}_F$$

According to the temperature definition in the thermodynamics (Annamalai et al., 2011) and comparing Eq. (37) with Eq. (35), there are:

$$\theta = \frac{\partial \xi_S}{\delta \eta_S} = \frac{\partial \xi_F}{\delta \eta_F}; \tilde{\boldsymbol{T}}_H = \rho_S \frac{\partial \xi_S}{\partial \boldsymbol{E}_H} + \rho_F \frac{\partial \xi_F}{\partial \boldsymbol{E}_H}; P_S = \rho_{RS} \frac{\partial \xi_S}{\partial \vartheta_S}; P_F = \rho_{RF} \frac{\partial \xi_F}{\partial \vartheta_F}. \quad (38a)$$

According to the thermodynamic theory (Annamalai et al., 2011; Bennethum, 1996) and internal variable theory (Collin & Houlsby,1997), we have:

$$\bar{\mu}_F^L = \frac{\partial \xi_F}{\partial c_F^L}; \bar{\mu}_F^c = \frac{\partial \xi_F}{\partial c_F^c}; \tilde{T}_H^p = -\rho_S \frac{\partial \xi_S}{\partial E_H^p} - \rho_F \frac{\partial \xi_F}{\partial E_H^p}. \tag{38b}$$

where $\bar{\mu}_F^L, \bar{\mu}_F^c$ are the chemical potential with respects to $c_F^L$ and $c_F^c$ in the fluid phase, respectively. According to the thermodynamic theory, $\bar{\mu}_F^\beta = \mu_F^\beta / M_F^\beta$, $M_F^\beta$ is the molar mass of $\beta$-constituent in the fluid phase; $\mu_F^\beta$ is the chemical potential of $\beta$-constituent in the unit molar of fluid phase; $\tilde{T}_H^p$ is the Kirchhoff-type dissipative Terzaghi effective stress. Noting that Eq. (38b)$_3$ only reflect the combined relationship between the elastic and plastic mechanic element (Collin & Houlsby,1997), Eqs. (38a) and (38b) are actually the general elastic constitutive equations of saturated porous media in terms of internal energy potential function.

Let $\hat{\boldsymbol{p}}_A = \rho_F \frac{\partial \xi_F}{\partial \boldsymbol{E}_H} : \nabla \boldsymbol{E}_H + \rho_F \frac{\partial \xi_F}{\partial \boldsymbol{E}_H^p} : \nabla \boldsymbol{E}_H^p$ and $\psi_\alpha = \xi_\alpha - \theta \eta_\alpha$, Substituting Eq. (37) into Eq. (35) and using Eqs. (38a) and (38b) gives (De Groot & Mazur, 1962):

$$\theta \sum_{\alpha=S,F} \rho_\alpha \frac{d^\alpha \eta_\alpha}{dt} = -\rho_F \bar{\mu}_F^L \frac{d^F c_F^L}{dt} - \rho_F \bar{\mu}_F^c \frac{d^F c_F^c}{dt} + \tilde{T}_H^p \frac{d^S \boldsymbol{E}_H^p}{dt} + \sum_{\alpha=S,F}(-\nabla \cdot \boldsymbol{q}_\alpha + \rho_\alpha r_\alpha) \\ +(P_F \nabla \varphi_F - \hat{\boldsymbol{p}}_F + \hat{\boldsymbol{p}}_A) \cdot \boldsymbol{W}_F - \sum_{\alpha=S,F} \hat{c}_\alpha \left( \frac{P_\alpha}{\rho_{R\alpha}} + \frac{1}{2} W_\alpha^2 + \psi_\alpha + \theta \eta_\alpha \right) + \sum_{\beta=L,c} \rho_F^\beta \boldsymbol{u}_F^\beta \cdot \boldsymbol{b}_F^\beta \tag{39}$$

Noting that the following identity equation:

$$\rho \frac{d\eta}{dt} = \sum_{\alpha=S,F} \left[ \rho_\alpha \frac{d^\alpha \eta_\alpha}{dt} - \nabla \cdot (\rho_\alpha \eta_\alpha \boldsymbol{u}_\alpha) + \hat{c}_\alpha \eta_\alpha \right] \tag{40}$$

Let $\bar{\mu}_F = c_F^L \bar{\mu}_F^L + c_F^c \bar{\mu}_F^c$, substituting Eqs.(16) and (40) into Eq.(39) gives

$$\rho \frac{\mathrm{d}\eta}{\mathrm{d}t} = -\sum_{\alpha=S,F} \nabla \cdot (\rho_\alpha \eta_\alpha \boldsymbol{u}_\alpha) + \sum_{\beta=L,c} \frac{\bar{\mu}_F^\beta}{\theta} \nabla \cdot (\rho_F^\beta \boldsymbol{u}_F^\beta) + \frac{\tilde{\boldsymbol{T}}_H^p}{\theta} \frac{\mathrm{d}^S \boldsymbol{E}_H^p}{\mathrm{d}t} -$$
$$\sum_{\alpha=S,F} \frac{1}{\theta} \nabla \cdot \boldsymbol{q}_\alpha + \sum_{\alpha=S,F} \frac{\rho_\alpha r_\alpha}{\theta} + \frac{P_F \nabla \varphi_F - \hat{\boldsymbol{p}}_F + \hat{\boldsymbol{p}}_A}{\theta} \cdot \boldsymbol{W}_F - \sum_{\beta=L,c} \frac{\bar{\mu}_F^\beta \hat{c}_F^\beta}{\theta} - \qquad (41)$$
$$\sum_{\alpha=S,F} \frac{\hat{c}_\alpha}{\theta} \left( \psi_\alpha + \frac{1}{2} W_\alpha^2 + \frac{P_\alpha}{\rho_{R\alpha}} \right) + \frac{\hat{c}_F \bar{\mu}_F}{\theta} + \sum_{\beta=L,c} \frac{\rho_F^\beta \boldsymbol{u}_F^\beta}{\theta} \cdot \boldsymbol{b}_F^\beta$$

Using the identity equation $b\nabla \cdot \boldsymbol{a} = \nabla \cdot (b\boldsymbol{a}) - \boldsymbol{a} \cdot \nabla b$, Eq.(41) can be rewritten as follows:

$$\rho \frac{\mathrm{d}\eta}{\mathrm{d}t} = \nabla \cdot [-\sum_{\alpha=S,F} (\rho_\alpha \eta_\alpha \boldsymbol{u}_\alpha + \frac{\boldsymbol{q}_\alpha}{\theta}) + \sum_{\beta=L,c} \frac{1}{\theta} \bar{\mu}_F^\beta \rho_F^\beta \boldsymbol{u}_F^\beta] + \sum_{\alpha=S,F} \frac{\rho_\alpha r_\alpha}{\theta} + \frac{\tilde{\boldsymbol{T}}_H^p}{\theta} \frac{\mathrm{d}^S \boldsymbol{E}_H^p}{\mathrm{d}t}$$
$$+ \frac{P_F \nabla \varphi_F - \hat{\boldsymbol{p}}_F + \hat{\boldsymbol{p}}_A}{\theta} \cdot \boldsymbol{W}_F - \sum_{\beta=L,c} \rho_F^\beta \boldsymbol{u}_F^\beta \cdot \nabla(\frac{\bar{\mu}_F^\beta}{\theta}) + \sum_{\alpha=S,F} \boldsymbol{q}_\alpha \cdot \nabla \frac{1}{\theta} \qquad (42)$$
$$- \sum_{\alpha=S,F} \frac{\hat{c}_\alpha}{\theta} \left( \psi_\alpha + \frac{1}{2} W_\alpha^2 + \frac{P_\alpha}{\rho_{R\alpha}} \right) + \frac{\hat{c}_F \bar{\mu}_F}{\theta} - \sum_{\beta=L,c} \frac{\bar{\mu}_F^\beta \hat{c}_F^\beta}{\theta} + \sum_{\beta=L,c} \frac{\rho_F^\beta \boldsymbol{u}_F^\beta}{\theta} \cdot \boldsymbol{b}_F^\beta$$

According to non-equilibrium thermodynamics (De Groot & Mazur, 1962), the entropy inequality can be obtained from the energy balance equation and the free energy potential function. According to Eq. (42) and the entropy decomposition theory of non-equilibrium thermodynamics (De Groot & Mazur, 1962), the entropy flow can be expressed from Eq. (42) as:

$$\rho \frac{\mathrm{d}\eta^r}{\mathrm{d}t} = \nabla \cdot [-\sum_{\alpha=S,F} (\rho_\alpha \eta_\alpha \boldsymbol{u}_\alpha + \frac{\boldsymbol{q}_\alpha}{\theta}) + \sum_{\beta=L,c} \frac{1}{\theta} \bar{\mu}_F^\beta \rho_F^\beta \boldsymbol{u}_F^\beta] + \sum_{\alpha=S,F} \frac{\rho_\alpha r_\alpha}{\theta} \qquad (43)$$

The entropy production can be expressed from Eq. (42) as:

$$\rho \frac{\mathrm{d}\eta^i}{\mathrm{d}t} = \frac{\tilde{\boldsymbol{T}}_H^p}{\theta} \frac{\mathrm{d}^S \boldsymbol{E}_H^p}{\mathrm{d}t} + \frac{P_F \nabla \varphi_F - \hat{\boldsymbol{p}}_F + \hat{\boldsymbol{p}}_A}{\theta} \cdot \boldsymbol{W}_F - \sum_{\beta=L,c} \rho_F^\beta \boldsymbol{u}_F^\beta \cdot [\nabla(\frac{\bar{\mu}_F^\beta}{\theta}) - \frac{\boldsymbol{b}_F^\beta}{\theta}]$$
$$+ \sum_{\alpha=S,F} \boldsymbol{q}_\alpha \cdot \nabla \frac{1}{\theta} - \sum_{\alpha=S,F} \frac{\hat{c}_\alpha}{\theta} \left( \psi_\alpha + \frac{1}{2} W_\alpha^2 + \frac{P_\alpha}{\rho_{R\alpha}} \right) + \frac{\hat{c}_F \bar{\mu}_F}{\theta} - \sum_{\beta=L,c} \frac{\bar{\mu}_F^\beta \hat{c}_F^\beta}{\theta} \qquad (44)$$

On the right side of Eq. (44), the first item is the entropy production due to the plastic behavior. The second item is the entropy production due to seepage. The third item is the entropy production due to diffusion in the fluid. The forth item is the entropy production due to heat flux. The fifth and sixth items are the entropy production due to the mass exchange of chemical action between the solid and the fluid. The final item is the entropy production due to the mass exchange of chemical between the constituents in the fluid phase. According to the second law of thermodynamics (Annamalai et al., 2011), the entropy production, i.e. Eq.(44), should be greater than or equal to zero. Eqs.(10) and (11) yield $\rho_F^L u_F^L = -\rho_F^c u_F^c$, so the following dissipation function can be derived from Eq. (44), $\rho_F^L u_F^L = -\rho_F^c u_F^c$ and the second law of thermodynamics:

$$\varsigma = \rho\theta \frac{d\eta^i}{dt} = \tilde{T}_H^p : \frac{d^S E_H^p}{dt} + (P_F \nabla \varphi_F - \hat{p}_F + \hat{p}_A) \cdot W_F$$
$$-\rho_F^c u_F^c \cdot [\theta \nabla (\frac{\bar{\mu}_F^c - \bar{\mu}_F^L}{\theta}) - b_F^c + b_F^L] + \sum_{\alpha=S,F} q_\alpha \cdot \theta \nabla \frac{1}{\theta} \quad (45)$$
$$-\sum_{\alpha=S,F} \hat{c}_\alpha \left( \psi_\alpha + \frac{1}{2} W_\alpha^2 + \frac{P_\alpha}{\rho_{R\alpha}} \right) + \hat{c}_F \bar{\mu}_F - \sum_{\beta=L,c} \bar{\mu}_F^\beta \hat{c}_F^\beta \geq 0$$

**5. General constitutive equation under infinitesimal strain condition**

Under infinitesimal strain condition, the higher order items can be omitted. The spatial derivative is equal to the material derivative. Let $U_S$ be the displacement of the solid, i.e. $U_S = x_S - X_S$, $\varepsilon_S$ and $\varepsilon_S^p$ denote the solid strain tensor and the solid plastic strain tensor under infinitesimal strain condition, $\varepsilon_{SV}$ and $\varepsilon_{SV}^p$ are the corresponding solid volumetric strain and plastic volumetric strain of the solid. $U_H$ is the displacement of the solid skeleton, i.e. $U_H = x_S - x_M$, $x_M$ is the position after matrix (material) deformation and mass exchange deformation, $\varepsilon_H$ and $\varepsilon_H^p$ denote the strain tensor of the solid skeleton and the plastic strain tensor of the solid skeleton under in-finitesimal strain condition, $\varepsilon_{HV}$ and $\varepsilon_{HV}^p$ are the corresponding volumetric strain of the solid skeleton and plastic volumetric strain of the solid skeleton. Under infinitesimal strain condition, there are $\boldsymbol{D}_S = \mathrm{d}\boldsymbol{\varepsilon}_S/\mathrm{d}t$, $\boldsymbol{D}_H = \mathrm{d}\boldsymbol{\varepsilon}_H/\mathrm{d}t$, $\rho_S \approx \rho_{S0}$, $\rho_{RS} \approx \rho_{RS0}$, $\rho_F \approx \rho_{F0}$, $\rho_{RF} \approx \rho_{RF0}$, $\tilde{\boldsymbol{T}}_H \approx \tilde{\boldsymbol{\sigma}}$ and $\tilde{\boldsymbol{T}}_H^p \approx \tilde{\boldsymbol{\sigma}}^p$. Based on the definition of $\boldsymbol{E}_S$ and $\boldsymbol{E}_H$, we have

$$\begin{aligned}\boldsymbol{E}_S &= \frac{1}{2}\left[\left(\frac{\partial \boldsymbol{U}_H^T}{\partial \boldsymbol{X}_S}+\boldsymbol{I}\right)\left(\frac{\partial \boldsymbol{U}_H}{\partial \boldsymbol{X}_S}+\boldsymbol{I}\right)-\boldsymbol{I}\right]-\vartheta_S\frac{\boldsymbol{I}}{3}+\varepsilon_c\frac{\boldsymbol{I}}{3}\\ &= \frac{1}{2}\left(\frac{\partial \boldsymbol{U}_H^T}{\partial \boldsymbol{X}_S}+\frac{\partial \boldsymbol{U}_H}{\partial \boldsymbol{X}_S}+\frac{\partial \boldsymbol{U}_H^T}{\partial \boldsymbol{X}_S}\frac{\partial \boldsymbol{U}_H}{\partial \boldsymbol{X}_S}\right)-\vartheta_S\frac{\boldsymbol{I}}{3}+\varepsilon_c\frac{\boldsymbol{I}}{3}\end{aligned} \quad (46)$$

$$\boldsymbol{E}_H = \frac{1}{2}\left[\left(\frac{\partial \boldsymbol{U}_H^T}{\partial \boldsymbol{X}_S}+\boldsymbol{I}\right)\left(\frac{\partial \boldsymbol{U}_H}{\partial \boldsymbol{X}_S}+\boldsymbol{I}\right)-\boldsymbol{I}\right] = \frac{1}{2}\left(\frac{\partial \boldsymbol{U}_H^T}{\partial \boldsymbol{X}_S}+\frac{\partial \boldsymbol{U}_H}{\partial \boldsymbol{X}_S}+\frac{\partial \boldsymbol{U}_H^T}{\partial \boldsymbol{X}_S}\frac{\partial \boldsymbol{U}_H}{\partial \boldsymbol{X}_S}\right) \quad (47)$$

The highest order items in Eqs.(46) and (47) are omitted under infinitesimal strain condition and noting that $\boldsymbol{\varepsilon}_S$ and $\boldsymbol{\varepsilon}_H$ are the linear part of $\boldsymbol{E}_S$ and $\boldsymbol{E}_H$, respectively, we have:

$$\boldsymbol{\varepsilon}_S = \frac{1}{2}\left(\frac{\partial \boldsymbol{U}_H^T}{\partial \boldsymbol{X}_S}+\frac{\partial \boldsymbol{U}_H}{\partial \boldsymbol{X}_S}\right)-\vartheta_S\frac{\boldsymbol{I}}{3}+\varepsilon_c\frac{\boldsymbol{I}}{3} \quad (48)$$

$$\boldsymbol{\varepsilon}_H = \frac{1}{2}\left(\frac{\partial \boldsymbol{U}_H^T}{\partial \boldsymbol{X}_S}+\frac{\partial \boldsymbol{U}_H}{\partial \boldsymbol{X}_S}\right) \quad (49)$$

Since $\mathrm{d}\varepsilon_{HV}/\mathrm{d}t = \mathrm{tr}\,\boldsymbol{D}_H = -\mathrm{d}^S\varphi_S/(\varphi_S\mathrm{d}t) = \mathrm{d}^S\ln(\varphi_{S0}/\varphi_S)/\mathrm{d}t$ which can derived from Eqs. (30) and (32) under infinitesimal strain condition, $\varepsilon_{HV} = \ln(\varphi_{S0}/\varphi_S)$. Under infinitesimal strain condition, the higher order items of Taylor expansion of $\ln(\varphi_{S0}/\varphi_S)$ can be omitted, we have

$$\varepsilon_{HV} = \frac{\varphi_{S0}-\varphi_S}{\varphi_{S0}} \quad (50)$$

where $\varphi_{S0}$ is the initial volume fraction of the solid. Base on the same reason, from the definition of the matrix strain $\vartheta_\alpha = \ln(\rho_{R\alpha}/\rho_{R\alpha 0})$, we can obtained under infinitesimal strain condition:

$$\vartheta_\alpha = \frac{\rho_{R\alpha} - \rho_{R\alpha 0}}{\rho_{R\alpha 0}} \tag{51}$$

According to $\boldsymbol{D}_S = \mathrm{d}\boldsymbol{\varepsilon}_S/\mathrm{d}t$, $\boldsymbol{D}_H = \mathrm{d}\boldsymbol{\varepsilon}_H/\mathrm{d}t$ and Eq. (30), under infinitesimal strain condition we can obtain (Noting that $\boldsymbol{\varepsilon}_S$ and $\boldsymbol{\varepsilon}_H$ take the tension as positive while $\vartheta_\alpha$ takes the compression as positive):

$$\boldsymbol{\varepsilon}_S = \boldsymbol{\varepsilon}_H - \frac{1}{3}\vartheta_S \boldsymbol{I} + \frac{1}{3}\varepsilon_c \boldsymbol{I} \tag{52}$$

and

$$\varepsilon_{SV} = \varepsilon_{HV} - \vartheta_S + \varepsilon_c \tag{53}$$

From Eq. (53), it can be seen that under infinitesimal strain condition, the volumetric strain of the solid phase can be divided into the solid skeleton volumetric strain, the solid matrix volumetric strain and the mass-exchange strain.

Under infinitesimal strain condition, Eqs. (38a) and (38b) become:

$$\left.\begin{aligned}
\theta &= \frac{\partial \xi_S(\eta_S, \boldsymbol{\varepsilon}_H, \boldsymbol{\varepsilon}_H^p, \vartheta_S)}{\partial \eta_S} = \frac{\partial \xi_F(\eta_F, \boldsymbol{\varepsilon}_H, \boldsymbol{\varepsilon}_H^p, \vartheta_F, c_F^L, c_F^c)}{\partial \eta_F}; \\
\tilde{\boldsymbol{\sigma}} &= \frac{\partial[\rho_{S0}\xi_S(\eta_S, \boldsymbol{\varepsilon}_H, \boldsymbol{\varepsilon}_H^p, \vartheta_S) + \rho_{F0}\xi_F(\eta_F, \boldsymbol{\varepsilon}_H, \boldsymbol{\varepsilon}_H^p, \vartheta_F, c_F^L, c_F^c)]}{\partial \boldsymbol{\varepsilon}_H}; \\
\tilde{\boldsymbol{\sigma}}^p &= -\frac{\partial[\rho_{S0}\xi_S(\eta_S, \boldsymbol{\varepsilon}_H, \boldsymbol{\varepsilon}_H^p, \vartheta_S) + \rho_{F0}\xi_F(\eta_F, \boldsymbol{\varepsilon}_H, \boldsymbol{\varepsilon}_H^p, \vartheta_F, c_F^L, c_F^c)]}{\partial \boldsymbol{\varepsilon}_H^p}; \\
P_S &= \frac{\partial[\rho_{RS0}\xi_S(\eta_S, \boldsymbol{\varepsilon}_H, \boldsymbol{\varepsilon}_H^p, \vartheta_S)]}{\partial \vartheta_S}; P_F = \frac{\partial[\rho_{RF0}\xi_F(\eta_F, \boldsymbol{\varepsilon}_H, \boldsymbol{\varepsilon}_H^p, \vartheta_F, c_F^L, c_F^c)]}{\partial \vartheta_F}; \\
\overline{\mu}_F^L &= \frac{\partial \xi_F(\eta_F, \boldsymbol{\varepsilon}_H, \boldsymbol{\varepsilon}_H^p, \vartheta_F, c_F^L, c_F^c)}{\partial c_F^L}; \overline{\mu}_F^c = \frac{\partial \xi_F(\eta_F, \boldsymbol{\varepsilon}_H, \boldsymbol{\varepsilon}_H^p, \vartheta_F, c_F^L, c_F^c)}{\partial c_F^c}
\end{aligned}\right\} \tag{54}$$

where $\tilde{\boldsymbol{\sigma}}^p$ is a dissipative stress (Collin & Houlsby,1997). Helmholtz free energy $\rho_{\alpha 0}\psi_\alpha = \rho_{\alpha 0}\xi_\alpha - \theta\rho_{\alpha 0}\eta_\alpha$ is introduced, $\alpha \in \{S, F\}$. Substituting Eq. (54) into the

complete differentiation of Helmholtz free energy yields:

$$\left.\begin{array}{l} \eta_S = \dfrac{\partial \psi_S(\theta,\varepsilon_H,\varepsilon_H^p,\vartheta_S)}{\partial \theta}; \eta_F = \dfrac{\partial \psi_F(\theta,\varepsilon_H,\varepsilon_H^p,\vartheta_F,c_F^L,c_F^c)}{\partial \theta}; \\[6pt] \tilde{\sigma} = \dfrac{\partial[\rho_{S0}\psi_S(\theta,\varepsilon_H,\varepsilon_H^p,\vartheta_S) + \rho_{F0}\psi_F(\theta,\varepsilon_H,\varepsilon_H^p,\vartheta_F,c_F^L,c_F^c)]}{\partial \varepsilon_H}; \\[6pt] \tilde{\sigma}^p = -\dfrac{\partial[\rho_{S0}\psi_S(\theta,\varepsilon_H,\varepsilon_H^p,\vartheta_S) + \rho_{F0}\psi_F(\theta,\varepsilon_H,\varepsilon_H^p,\vartheta_F,c_F^L,c_F^c)]}{\partial \varepsilon_H^p} \\[6pt] P_S = \dfrac{\partial[\rho_{RS0}\psi_S(\theta,\varepsilon_H,\varepsilon_H^p,\vartheta_S)]}{\partial \vartheta_S}; P_F = \dfrac{\partial[\rho_{RF0}\psi_F(\theta,\varepsilon_H,\varepsilon_H^p,\vartheta_F,c_F^L,c_F^c)]}{\partial \vartheta_F} \\[6pt] \bar{\mu}_F^L = \dfrac{\partial \psi_F(\theta,\varepsilon_H,\varepsilon_H^p,\vartheta_F,c_F^L,c_F^c)}{\partial c_F^L}; \bar{\mu}_F^c = \dfrac{\partial \psi_F(\theta,\varepsilon_H,\varepsilon_H^p,\vartheta_F,c_F^L,c_F^c)}{\partial c_F^c} \end{array}\right\} \quad (55)$$

Because the temperature $\theta$ of the solid and fluid phases are assumed to be constant, $\theta$ in Eq. (55) can be ignored without writing in the next text and Eq. (55)$_{1\text{-}1}$ and Eq. (55)$_{1\text{-}2}$ can also be omitted. Under most cases for practical purpose, the Helmholtz free energy $\psi_S(\varepsilon_H,\varepsilon_H^p,\vartheta_S)$ of the solid phase can be simplified as two parts: (i) one is a free energy of $\psi_{RS}(\vartheta_S)$ due to the volumetric deformation of the solid matrix; (ii) another is a free energy of $\psi_{SH}(\varepsilon_H,\varepsilon_H^p)$ due to the deformation of the solid skeleton. Due to $c_F^L = 1 - c_F^c$, the Helmholtz free energy $\psi_F(\varepsilon_{Sf},\varepsilon_{Sf}^p,\vartheta_F,c_F^L,c_F^c)$ of the fluid phase can also be simplified as $\psi_F(\varepsilon_{Sf},\varepsilon_{Sf}^p,\vartheta_F,c_F^c)$. For most cases, $\psi_F(\varepsilon_H,\varepsilon_H^p,\vartheta_F,c_F^c)$ is also simplified as two parts: (i) one is a free energy of $\psi_{RF}(\vartheta_F,c_F^c)$ due to the volumetric deformation and the chemical activity of the fluid matrix; (ii) another is a free energy of $\psi_{FH}(\varepsilon_H,\varepsilon_H^p)$ due to the absorb action expressed by the skeleton strain (Bennethum, et al, 2000). Let $\Psi_{HS}(\varepsilon_H,\varepsilon_H^p) = \rho_{S0}\psi_{HS}(\varepsilon_H,\varepsilon_H^p)$, $\Psi_{HF}(\varepsilon_H,\varepsilon_H^p) = \rho_{F0}\psi_{HF}(\varepsilon_H,\varepsilon_H^p)$, $\Psi_{RS}(\vartheta_S) = \rho_{RS0}\psi_{RS}(\vartheta_S)$, $\Psi_{RF}(\vartheta_F,c_F^c) = \rho_{RF0}\psi_{RF}(\vartheta_F,c_F^c)$, using the above relationship and equations, Eq.(55) becomes:

$$\left.\begin{aligned}
\tilde{\boldsymbol{\sigma}} &= \frac{\partial[\Psi_{HS}(\boldsymbol{\varepsilon}_H, \boldsymbol{\varepsilon}_H^p) + \Psi_{HF}(\boldsymbol{\varepsilon}_H, \boldsymbol{\varepsilon}_H^p)]}{\partial \boldsymbol{\varepsilon}_H}; \\
\tilde{\boldsymbol{\sigma}}^p &= -\frac{\partial[\Psi_{HS}(\boldsymbol{\varepsilon}_H, \boldsymbol{\varepsilon}_H^p) + \Psi_{HF}(\boldsymbol{\varepsilon}_H, \boldsymbol{\varepsilon}_H^p)]}{\partial \boldsymbol{\varepsilon}_H^p} \\
P_S &= \frac{\partial \Psi_{RS}(\vartheta_S)}{\partial \vartheta_S}; P_F = \frac{\partial \Psi_{RF}(\vartheta_F, c_F^c)}{\partial \vartheta_F}; \\
\bar{\mu}_F^c - \bar{\mu}_F^L &= \frac{1}{\rho_{RF0}} \frac{\partial \Psi_{RF}(\vartheta_F, c_F^c)}{\partial c_F^c}
\end{aligned}\right\} \quad (56)$$

Eq.(56) is the general elastic constitutive equation in terms of the free potential function under infinitesimal strain condition. Because the temperature $\theta$ of the solid and fluid phases remains a constant, the dissipation function Eq. (45) under infinitesimal strain condition becomes:

$$\begin{aligned}
\varsigma = \rho\theta \frac{d\eta^i}{dt} &= \tilde{\boldsymbol{\sigma}}^p : \frac{d\boldsymbol{\varepsilon}_H^p}{dt} + (P_F \nabla \varphi_F - \hat{\boldsymbol{p}}_F + \hat{\boldsymbol{p}}_A) \cdot \boldsymbol{W}_F - \\
& \rho_F^c \boldsymbol{u}_F^c \cdot [\nabla(\bar{\mu}_F^c - \bar{\mu}_F^L) - \boldsymbol{b}_F^c + \boldsymbol{b}_F^L] + \hat{c}_F \bar{\mu}_F \\
& - \sum_{\alpha=S,F} \hat{c}_\alpha \left(\psi_\alpha + \frac{1}{2}\boldsymbol{W}_\alpha^2 + \frac{P_\alpha}{\rho_{R\alpha}}\right) - \sum_{\beta=L,c} \bar{\mu}_F^\beta \hat{c}_F^\beta \geq 0
\end{aligned} \quad (57)$$

When the following inequalities hold, Eq. (57) is definitely valid.

$$\tilde{\boldsymbol{\sigma}}^p : \frac{d\boldsymbol{\varepsilon}_H^p}{dt} \geq 0, \quad (P_F \nabla \varphi_F - \hat{\boldsymbol{p}}_F + \hat{\boldsymbol{p}}_A) \cdot \boldsymbol{W}_F \geq 0; \quad (58a)$$

$$-\rho_F^c \boldsymbol{u}_F^c \cdot [\nabla(\bar{\mu}_F^c - \bar{\mu}_F^L) - \boldsymbol{b}_F^c + \boldsymbol{b}_F^L] \geq 0; \quad (58b)$$

$$\begin{aligned}
& \hat{c}_F \bar{\mu}_F - \sum_{\alpha=S,F} \hat{c}_\alpha \left(\psi_\alpha + \frac{1}{2}\boldsymbol{W}_\alpha^2 + \frac{P_\alpha}{\rho_{R\alpha}}\right) - \sum_{\beta=L,c} \bar{\mu}_F^\beta \hat{c}_F^\beta \\
& = -\hat{c}_F \left(\psi_F + \frac{1}{2}\boldsymbol{W}_F^2 + \frac{P_F}{\rho_{RF}} - \bar{\mu}_F + \bar{\mu}_F^L - \psi_S - \frac{P_S}{\rho_{RS}}\right) - (\bar{\mu}_F^c - \bar{\mu}_F^L)\hat{c}_F^c \geq 0
\end{aligned} \quad (58c)$$

$\hat{c}_S + \hat{c}_F = 0$ and $\hat{c}_F^L + \hat{c}_F^c = \hat{c}_F$ are used in the derivation of Eq.(58c). Similar to Collins and Houlsby (1997), the dissipative potentials is generally assumed to exist so that:

$$\tilde{\boldsymbol{\sigma}}^p = \frac{\partial \Theta_p(d\boldsymbol{\varepsilon}_H^p / dt)}{\partial(d\boldsymbol{\varepsilon}_H^p / dt)}; P_F \nabla \varphi_F - \hat{\boldsymbol{p}}_F + \hat{\boldsymbol{p}}_A = \frac{\partial \Theta_w(\boldsymbol{W}_F)}{\partial \boldsymbol{W}_F}; \quad (59a)$$

$$\nabla(\overline{\mu}_F^c - \overline{\mu}_F^L) - \boldsymbol{b}_F^c + \boldsymbol{b}_F^L = -\frac{\partial \Theta_d(\rho_F^c \boldsymbol{u}_F^c)}{\partial(\rho_F^c \boldsymbol{u}_F^c)};  \quad (59b)$$

$$\psi_F + \frac{1}{2}W_F^2 + \frac{P_F}{\rho_{RF}} - \overline{\mu}_F + \overline{\mu}_F^L - \psi_S - \frac{P_S}{\rho_{RS}} = -\frac{\partial \Theta_c(\hat{c}_F)}{\partial \hat{c}_F}. \quad (59c)$$

The constitutive equation for inelastic physical and irreversible chemical processes can be obtained from Eqs. (59a) ~ (59c).

**6. Example**

Bentonite is an important buffer material in nuclear waste disposal repository, which is used to block the migration of high-level radioactive wastes such as nuclear waste. It is often subjected to the coupling action between chemical activity and mechanic force (Zhang et al., 2020). Many experiments have been performed on Bentonite and plenty of deformation characteristics and empirical formulas have been obtained from experiments (Di Maio et al., 2004; Ye et al., 2014; Zhang et al., 2016). The confined compression test of bentonite is a very important experiment. It can provide a key basis for establishing a constitutive model. In order to characterize the effect of chemical activity on mechanical properties and deformation properties of saturated bentonite, Zhang et al (2016) carried out a confined compression test on bentonite saturated with NaCl solution. In this confined compression test, bentonite was taken from the Inner Mongolia Autonomous Region, 300 km northwest from Beijing, China. It is a light gray powder, dominated by montmorillonite (75.4 % in mass). The liquid limit and plastic limit of the bentonite are 276 and 37 %, respectively (Zhang et al, 2016). The specific gravity is 2.66 and the dry density is 1.70 Mg/m$^3$. First, under the constant vertical stress of 0.5 MPa, the sample was hydrated to swell using either de-ionized water or NaCl solutions with concentrations of 0.5, 1 M, and 2 M. Meanwhile, the vertical displacement of the sample was recorded. Then, when the swelling strain reached a steady state, compression tests were conducted in a conventional way by successive loading to a maximum value of about 42 MPa. Noting that the mass fraction is adopted as chemical state variable in the constitutive framework in this paper, the concentration is transformed to the mass fraction, and the experimental data are processed and integrated, as shown in Fig 2~3. Now, we use this experimental data and the above-established constitutive theoretical framework considering chemical activity to capture the one-dimensional constitutive model of bentonite considering the chemo-hydro-mechanical coupling.

Under the confined compression, the horizontal solid strain is zero, i.e., only the vertical solid strain occurs. In order to simplify the analysis, the solid skeleton horizontal strain and the matrix horizontal strains are also assumed to be zero, i.e. only the vertical strains of the solid skeleton and matrix occur. Strictly speaking, this assumption does not hold, but when the strain of solid matrix is much smaller, it only brings little error. Under the above assumptions, the volumetric strain is equal to the vertical strain, $\Psi_{HS}(\varepsilon_H, \varepsilon_H^p)$ and $\Psi_{HF}(\varepsilon_H, \varepsilon_H^p)$ can be expressed as $\Psi_{HS}(\varepsilon_{Hz}, \varepsilon_{Hz}^p)$ and $\Psi_{HF}(\varepsilon_{Hz}, \varepsilon_{Hz}^p)$, where $\varepsilon_{Hz}$ and $\varepsilon_{Hz}^p$ are the vertical strain and the vertical plastic strain of the solid skeleton, respectively. The elastic constitutive equation in terms of the potential function can be obtained from Eq. (56):

$$\left.\begin{array}{l}\tilde{P}_z = -\dfrac{\partial[\Psi_{HS}(\varepsilon_{Hz},\varepsilon_{Hz}^p)+\Psi_{HF}(\varepsilon_{Hz},\varepsilon_{Hz}^p)]}{\partial \varepsilon_{Hz}} \\[2mm] \tilde{P}_z^p = \dfrac{\partial[\Psi_{HS}(\varepsilon_{Hz},\varepsilon_{Hz}^p)+\Psi_{HF}(\varepsilon_{Hz},\varepsilon_{Hz}^p)]}{\partial \varepsilon_{Hz}^p} \\[2mm] P_{Sz} = \dfrac{\partial \Psi_{RS}(\vartheta_{Sz})}{\partial \vartheta_{Sz}},\ P_F = \dfrac{\partial \Psi_{RF}(\vartheta_F, c_F^c)}{\partial \vartheta_F} \\[2mm] \bar{\mu}_F^c - \bar{\mu}_F^L = \dfrac{\partial \Psi_{RF}(\vartheta_F, c_F^c)}{\rho_{RF0}\partial c_F^c}\end{array}\right\} \quad (60)$$

where $\tilde{P}_z = -\sigma_z - P_F$ is Terzaghi's vertical effective pressure. $\tilde{P}_z = -\tilde{\sigma}_z$ due to $\tilde{\sigma}_z = \sigma_z + P_F$, Noting that $\tilde{\sigma}_z$ is positive in tension, so $\tilde{P}_z$ is positive in compression. $\tilde{P}_z^p = -\tilde{\sigma}_z^p$ is the dissipative vertical effective pressure. Similar to $\tilde{P}_z$, $\tilde{P}_z^p$ is also positive in compression. $P_{Sz}$ and $\vartheta_{Sz}$ are the vertical pressure and strain of the solid matrix, they are positive in compression. Eq. (60) includes the not only solid but also fluid constitutive equations, so a complete constitutive relationship should contain those of the solid and fluid phases. In addition, Eq. (60) still reveals that the mass fraction $c_F^c$ rather than the molar concentration in the NaCl solution is used as a state variable to establish the constitutive model.

For most elastoplastic problems, $\Psi_{HS}(\varepsilon_{Hz},\varepsilon_{Hz}^p)+\Psi_{HF}(\varepsilon_{Hz},\varepsilon_{Hz}^p)$ can be expressed as:

$$\Psi_{HS}(\varepsilon_{Hz},\varepsilon_{Hz}^p)+\Psi_{HF}(\varepsilon_{Hz},\varepsilon_{Hz}^p) = \Psi_{HS}(\varepsilon_{Hz}-\varepsilon_{Hz}^p)+\Psi_{HF}(\varepsilon_{Hz}-\varepsilon_{Hz}^p) \quad (61)$$

Substituting Eq. (61) into Eq. (60)$_2$ gives:

$$\tilde{P}_z^p = \tilde{P}_z \quad (62)$$

Eq. (62) indicates that the dissipative vertical pressure is equal to Terzaghi vertical effective pressure, so $\tilde{P}_z^p$ is also called Terzaghi vertical dissipative effective pressure.

Eq. (62) also indicates that Eq. (60)$_2$ yields the combination relationship of the elastic and plastic mechanic element and do not yield the specific plastic constitutive model.

In the current study of bentonite, both the mass exchange between the solid and fluid phases and the mass exchange between the constituents in the fluid phase are not considered. That is, $\hat{c}_F = \hat{c}_S = 0$ and $\hat{c}_F^L = \hat{c}_F^c = 0$, and $\varepsilon_c = 0$ can be derived from Eq. (28), so Eqs. (58b) and (59b) can be omitted. Under the confined compression condition, Eq. (52) becomes as:

$$\varepsilon_{Sz} = \varepsilon_{Hz} - \vartheta_{Sz} \tag{63}$$

where $\varepsilon_{Sz}$ is the vertical strain of the solid, $\varepsilon_{Hz}$ is the vertical strain of the solid skeleton, and $\vartheta_{Hz}$ is the vertical strain of the solid matrix. Based on Eq. (62) and the confined compression condition, Eqs. (58a) ~(58b) and (59a) ~ (59b) are degenerated into:

$$\left. \begin{array}{l} \tilde{P}_z \dfrac{d\varepsilon_{Hz}^p}{dt} \leq 0, \ (P_F \dfrac{\partial \varphi_F}{\partial z} - \hat{p}_{Fz} + \hat{p}_{Az})W_{Fz} \geq 0 \\ -\rho_F^c u_{Fz}^c \cdot [\dfrac{\partial(\bar{\mu}_F^c - \bar{\mu}_F^L)}{\partial z} - b_{Fz}^c + b_{Fz}^L] \geq 0 \end{array} \right\} ; \tag{64}$$

$$\left. \begin{array}{l} \tilde{P}_z = -\dfrac{\partial \Theta_p (d\varepsilon_{Hz}^p / dt)}{\partial (d\varepsilon_{Hz}^p / dt)}, \ P_F \dfrac{\partial \varphi_F}{\partial z} - \hat{p}_{Fz} + \hat{p}_{Az} = \dfrac{\partial \Theta_w(W_{Fz})}{\partial W_{Fz}} \\ \dfrac{\partial}{\partial z}(\bar{\mu}_F^c - \bar{\mu}_F^L) - b_{Fz}^c + b_{Fz}^L = -\dfrac{\partial \Theta_d(\rho_F^c u_{Fz}^c)}{\partial(\rho_F^c u_{Fz}^c)} \end{array} \right\} \tag{65}$$

where $z$ denotes the vertical coordinates. Eq. (65) shows that the constitutive relationship should include the plastic constitutive model and the seepage law of fluid as well as the diffusive equation of solute in the fluid. Next, under the guide of Eqs.(60) and (64) and (65), the complete constitutive relationships including the aspects of the solid and fluid phases as well as the solute in the fluid phase are built in the section.

6.1 Stress-strain equation of the solid matrix

The solid matrix strain is called the strain of soil particles or the solid material in soil mechanics. Since the deformation of the soil skeleton depends on the change of porosity, its strength is much smaller than that of the soil matrix. So the deformation of the solid matrix is within the elastic range before the soil skeleton is destroyed. Eq. (60)$_3$ indicates that the vertical strain of the solid matrix is uniquely determined by its vertical pressure. For a linear elastic model, Helmholtz free energy of the solid matrix can be expressed as

$$\Psi_{RS}(\vartheta_{Sz}) = \frac{1}{2} E_{RS} \vartheta_{Sz}^2 \tag{66}$$

where $E_{RS}$ is the compression modulus of the solid matrix under the confined compression condition. Substituting Eq. (66) into Eq. (60)$_3$ gives

$$P_{Sz} = E_{RS} \vartheta_{Sz} \tag{67}$$

6.2 Elastic stress-strain equation of the solid skeleton

Let the elastic strain of the solid skeleton be $\varepsilon_{Hz}^e = \varepsilon_{Hz} - \varepsilon_{Hz}^p$. Based on Eqs.(61) ~ (62) and the experimental data (Zhang et al 2016), the expression of $\Psi_{HS}(\varepsilon_{Hz}, \varepsilon_{Hz}^p)$ and $\Psi_{HF}(\varepsilon_{Hz}, \varepsilon_{Hz}^p)$ can be written as:

$$\left. \begin{array}{l} \Psi_{HS}(\varepsilon_{Hz}^e) = \tilde{P}_{z0} \lambda^e (\exp\dfrac{-\varepsilon_{Hz}^e}{\lambda^e} - 1) \\[2mm] \Psi_{HF}(\varepsilon_{Hz}^e) = \tilde{P}_{z0} \lambda^e [\exp\dfrac{\varepsilon_{Hmax} - \varepsilon_{Hz}^e}{\lambda^e} - \exp\dfrac{-\varepsilon_{Hz}^e}{\lambda^e} - \exp\dfrac{\varepsilon_{Hmax}}{\lambda^e} + 1] \end{array} \right\} \tag{68}$$

where $\tilde{P}_{z0}$ is the initial vertical pressure; $\lambda^e$ is the rebound index, the experiment shows that it remains constant along with the variation of the mass fraction of NaCl ($c_F^c$); $\varepsilon_{Hmax}$ is the free swelling strain at $\tilde{P}_z = \tilde{P}_{z0}$ for de-ionized water ($c_F^c = 0$). For saturated porous media, the expression between the volume fraction $\varphi_S$ of the solid phase and water content $\omega$ is $\varphi_S = d_F/(1+\omega d_S)$ (Lambe & Whitman, 1991), where $d_F$ and $d_S$ are the relative density of the fluid and solid matrices. Therefore, the relationship between the free swelling strain $\varepsilon_{Hmax}$ and the maximum swelling water content $\omega_{max}$ at $\tilde{P}_z = \tilde{P}_{z0}$ is $\varepsilon_{Hmax} = (\varphi_{S0} + \varphi_{S0}d_S\omega_{max} - d_F)/(\varphi_{S0} + \varphi_{S0}d_S\omega_{max})$. Substituting Eq. (68) into Eq. (60)$_1$ gives

$$\tilde{P}_z = \tilde{P}_{z0} \exp\frac{\varepsilon_{Hmax} - \varepsilon_{Hz}^e}{\lambda^e} \tag{69}$$

Eq. (69) can be transformed as:

$$\varepsilon_{Hz}^e = \varepsilon_{Hmax} - \lambda^e \ln\frac{\tilde{P}_z}{\tilde{P}_{z0}} \tag{70}$$

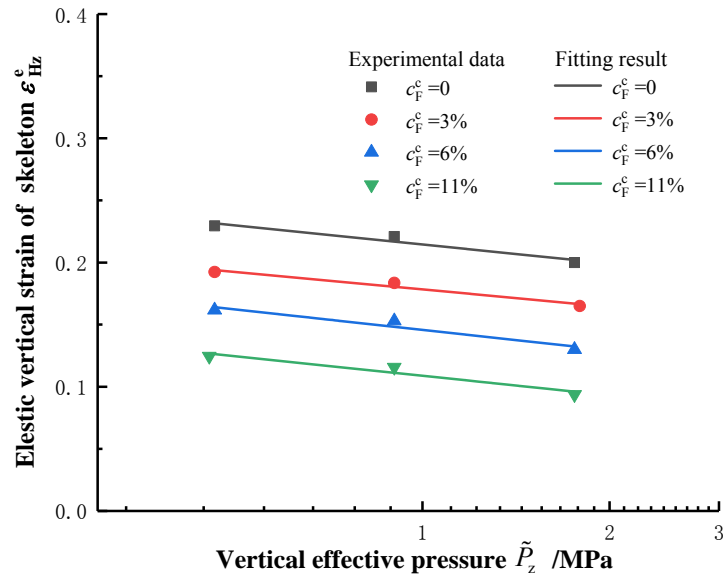

**Fig. 2** Variation of skeleton elastic vertical strain with vertical effective pressure

## 6.3 Plastic stress-strain equation of the solid skeleton

According to the experimental data and elastic stress-strain relationship, the plastic strain of bentonite saturated with different mass fractions of NaCl solutions can be obtained, which is shown in Fig.3 (Zhang et al., 2016). Combined with Eq. (65a)$_1$, the expression of $\Theta_p$ can be adopted as:

$$\Theta_p(\frac{d\varepsilon_{Hz}^p}{dt}) = -[\tilde{P}_{z0} \exp \frac{\varepsilon_{Hz0}^p(c_F^c) - \varepsilon_{Hz}^p}{\lambda^p(c_F^c)}]\frac{d\varepsilon_{Hz}^p}{dt} \tag{71}$$

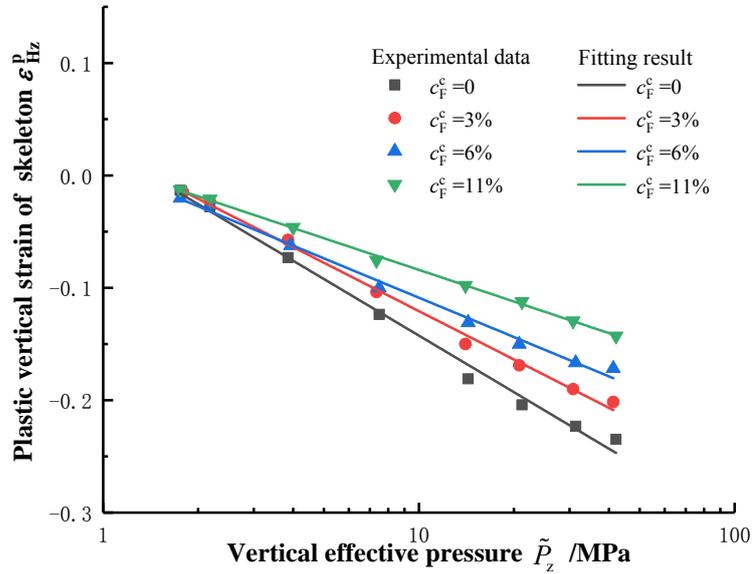

**Fig. 3** Variation of skeleton plastic vertical strain with vertical effective pressure

Substituting Eq.(71) into Eq.(65a)$_1$ gives

$$\tilde{P}_z = -\frac{\partial \Theta_p}{\partial(d\varepsilon_{Hz}^p / dt)} = \tilde{P}_{z0} \exp \frac{\varepsilon_{Hz0}^p(c_F^c) - \varepsilon_{Hz}^p}{\lambda^p(c_F^c)} \tag{72}$$

In order that Eq.(64)$_1$ is smaller than or equal to zero, $d\varepsilon_{Hz}^p / dt \leq 0$, i.e. $d\tilde{P}_z > 0$ is needed. Eq. (72) can be rewritten as:

$$\varepsilon_{Hz}^p = \varepsilon_{Hz0}^p(c_F^c) - \lambda^p(c_F^c) \ln \frac{\tilde{P}_z}{\tilde{P}_{z0}} \tag{73}$$

where $\tilde{P}_{z0}$ is the initial vertical effective pressure, $\lambda^p(c_F^c)$ is the plastic compression index. $\varepsilon_{Hz0}^{p}(c_F^c)$ is the initial vertical plastic strain at $\tilde{P}_z = \tilde{P}_{z0}$ on the compression line at the NaCl mass fraction $c_F^c$. The schematic diagram for Eq. (73) is shown in Fig.4.

The plastic compression index $\lambda^p(c_F^c)$ varies with $c_F^c$. The varying law can be determined by the following equation:

$$\lambda^p(c_F^c) = \lambda_1^p \exp(-c_F^c / \lambda_2^p) + \lambda_3^p \tag{74}$$

where $\lambda_1^p$, $\lambda_2^p$ and $\lambda_3^p$ are three fitting parameters. The relationship between Eq. (74) and the experimental data are shown in Fig 5 (Zhang et al., 2016), which indicates that the compressibility of bentonite decreases with the increase of the NaCl mass fraction.

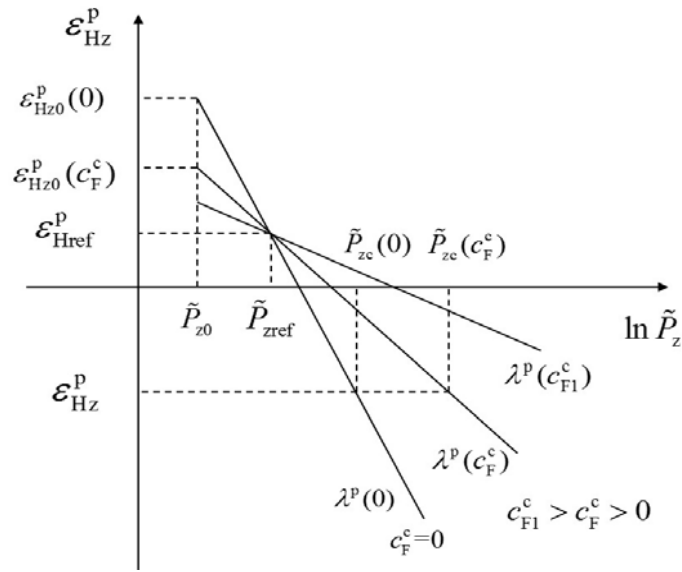

**Fig 4.** Plastic vertical stress-strain compression lines

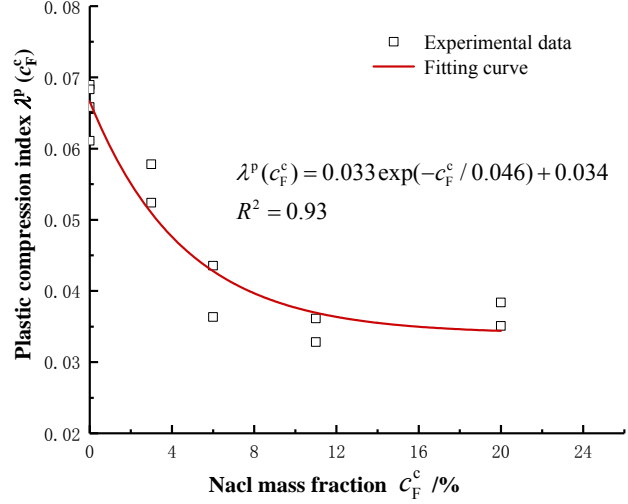

Fig 5. Relationship between the plastic compression index and NaCl mass fraction

In the classical soil mechanics, the vertical plastic strain of the solid skeleton $\varepsilon_{Hz}^p$ is chosen as a harden parameter, so $\varepsilon_{Hz}^p$ is constant on the same yielding surface. Similar to the suction effects on yield stress shown in Fig.4 (Alonso et al., 1990), it is assumed that all compression line pass through the same point which coordinates are $(\tilde{P}_{zref}, \varepsilon_{Href}^p)$ in the plane $\tilde{P}_z \sim \varepsilon_{Hz}^p$. Let the current yielding pressure at $c_F^c = 0$ for de-ionized water is denotes by $\tilde{P}_{zc}(0)$, the current yielding pressure at $c_F^c$ is denotes by $\tilde{P}_{zc}(c_F^c)$. $\tilde{P}_{z0}(c_F^c) = \tilde{P}_{zref}$ is adopted so that $\varepsilon_{Hz0}^p(c_F^c) = \varepsilon_{Href}^p$. Using Eq.(73) for different NaCl mass fractions, as shown in Fig.4, the effect of the NaCl mass fraction $c_F^c$ on the current yield pressure $\tilde{P}_{zc}(c_F^c)$ can be obtained as follows:

$$\varepsilon_{Hz}^p = \varepsilon_{Href}^p + \lambda^p(c_F^c)\ln[\tilde{P}_{zc}(c_F^c)/\tilde{P}_{zref}] = \varepsilon_{Href}^p + \lambda^p(0)\ln[\tilde{P}_{zc}(0)/\tilde{P}_{zref}] \qquad (75)$$

Simplifying Eq. (75) yields:

$$\tilde{P}_{zc}(c_F^c) = \tilde{P}_{zref}[\tilde{P}_{zc}(0)/\tilde{P}_{zref}]^{\frac{\lambda^p(0)}{\lambda^p(c_F^c)}} \qquad (76)$$

Comparison between the experimental data and the simulation results calculated from Eq. (76) are shown in Fig. 6 (Zhang et al., 2016), they are consistent with each other.

The experimental data also indicates that a plastic strain exists due to the change of NaCl mass fraction. This plastic strain is called the chemical plastic strain of the solid skeleton, which is independent of the dissipative potential and can be expressed as $\varepsilon_{Hz}^{pc}(c_F^c)$. Noting that $\tilde{P}_z$ is always $\tilde{P}_z > 0$, so $\tilde{P}_z \, d\varepsilon_{Hz}^{pc}(c_F^c)$ is always smaller than zero only if $d\varepsilon_{Hz}^{pc}(c_F^c) \leq 0$. That is, when $d\varepsilon_{Hz}^{pc}(c_F^c) \leq 0$, $\tilde{P}_z \, d\varepsilon_{Hz}^{pc}(c_F^c) \leq 0$ which can obey Eq. (64)$_1$. So introducing the chemical plastic strain does not violate the second law of thermodynamics. The experimental data describing the variation of $\varepsilon_{Hz}^{pc}(c_F^c)$ with $c_F^c$ are shown in Fig.7 (Zhang et al., 2016). Based on Fig.7, $\varepsilon_{Hz}^{pc}(c_F^c)$ can be expressed as:

$$\varepsilon_{Hz}^{pc}(c_F^c) = \varepsilon_p^c [\exp(-\lambda^c c_F^c) - 1] \tag{77}$$

where $\varepsilon_p^c$, $\lambda^c$ are chemical plastic strain parameters.

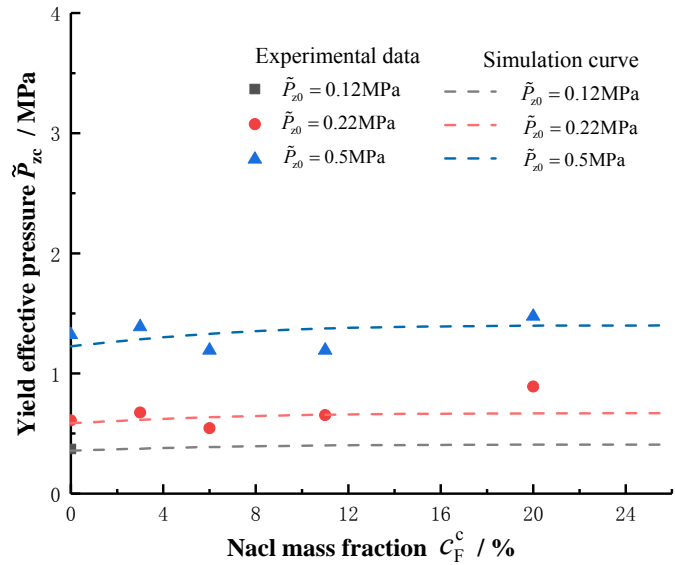

**Fig 6.** Comparison of simulation curves and test yield effective pressure

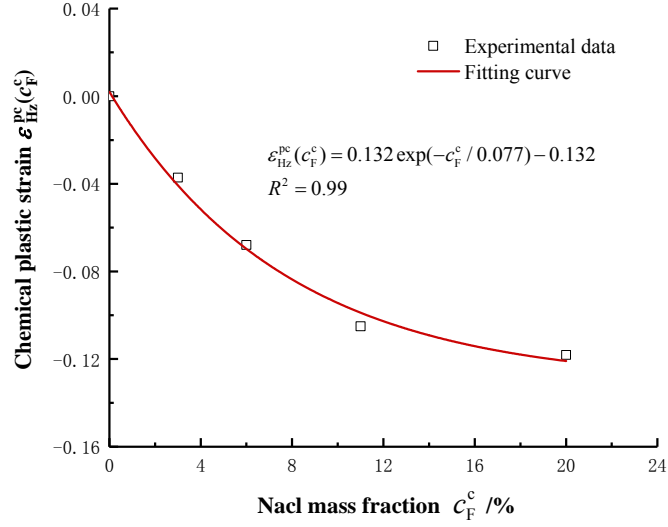

**Fig. 7** Variation of chemical plastic vertical strain with NaCl mass fraction

6.4 Stress-strain equation of the solid phase

Using Eqs. (63), (67), (70), (73), (77) and $\varepsilon_{Hz}^{e} = \varepsilon_{Hz} - \varepsilon_{Hz}^{p}$, and noting that $\varepsilon_{Hz}^{p}$ in fact include the plastic strain due to the effective pressure and the plastic strain due to the chemical activity, if the historic maximum yield pressure is $\tilde{P}_{zc}(c_F^c)$ at $c_F^c$ and the historic maximum mass fraction is $c_{Fc}^c$, we can uniformly get:

$$d\varepsilon_{Sz} = -\frac{\lambda}{\tilde{P}_z} d\tilde{P}_z - \varepsilon_p^c \lambda^c \exp(-\lambda^c c_F^c) dc_F^c - \frac{dP_S}{E_{RS}} \tag{78}$$

The parameters $\lambda^c$ and $\lambda$ in Eq. (78) can be adopted from Table.1.

Table.1 Adopting table of the parameters $\lambda^c$ and $\lambda$

| Case | $c_F^c < c_{Fc}^c$; $\tilde{P}_z < \tilde{P}_{zc}$ | $c_F^c \geq c_{Fc}^c$; $\tilde{P}_z < \tilde{P}_{zc}$ | $c_F^c \geq c_{Fc}^c$; $\tilde{P}_z \geq \tilde{P}_{zc}$ | $c_F^c < c_{Fc}^c$; $\tilde{P}_z \geq \tilde{P}_{zc}$ |
|---|---|---|---|---|
| $\lambda$ | $\lambda^e$ | $\lambda^e$ | $\lambda^e + \lambda^p(c_F^c)$ | $\lambda^e + \lambda^p(c_F^c)$ |
| $\lambda^c$ | 0 | $\lambda^c$ | $\lambda^c$ | 0 |

We use the above constitutive equations to simulate the experimental data. Fig 2~3

and Fig 5~7 provide the simulating results, and the parameters of the model determined by experimental data are shown in Table 2. The comparison of the simulation results predicted by these model parameters with the experimental data is drawn in Fig. 8 for the confined compression condition (Zhang et al., 2016). The comparing results that they are consistent with each other indicate that the HMT-based constitutive theoretical framework considering the porosity-dependent skeleton strain and the chemical activity can guide the establishment of the constitutive model of the solid phase for bentonite.

Combining the free swelling strain $\varepsilon_{Hmax}$ at ($\tilde{P}_z = \tilde{P}_{z0}$ and $c_F^c = 0$) with Eq. (78), we can obtain the current vertical swelling strain at any vertical effective pressure ($\tilde{P}_z$) and any NaCl mass fraction ($c_F^c$). If the current vertical strain is known, the relative vertical swelling strain quantity can be calculated from the difference between the current vertical swelling strain and the current vertical strain.

Table 2. Model parameters for confined compression tests

| $\lambda_1^p$ | $\lambda_2^p$ | $\lambda_3^p$ | $\lambda^c$ | $\lambda^e$ | $\varepsilon_p^c$ | $\varepsilon_{Hmax}$ | $\tilde{P}_{zref}$/MPa | $\tilde{P}_{zc}(0)$/MPa | $E_{RS}$/GPa |
|---|---|---|---|---|---|---|---|---|---|
| 0.033 | 0.046 | 0.034 | 1/0.077 | 0.02 | 0.132 | 0.23 | 1.04 | 1.23 | 20 |

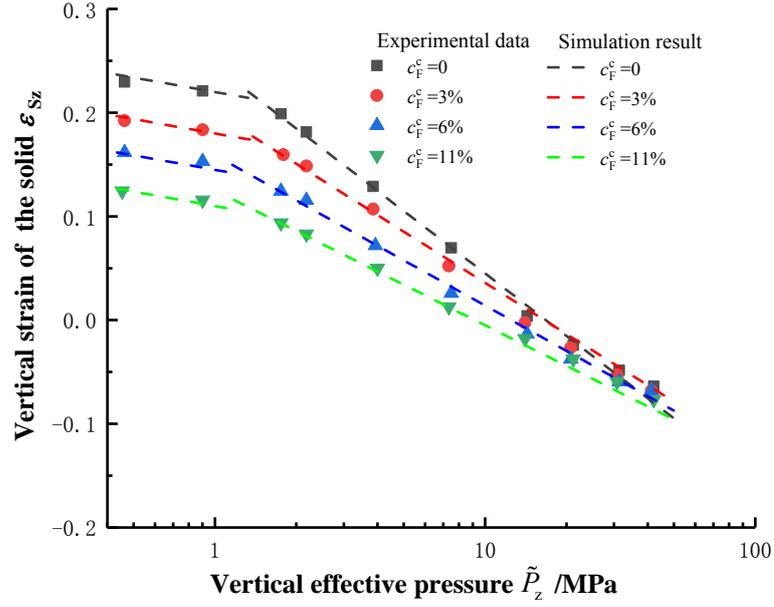

**Fig 8**. Comparison of experimental data and simulation results for confined compression tests

6.5 Stress-strain equation of the fluid matrix

Noting that the fluid matrix (i.e. the fluid in the pores) is the NaCl-water solution. According to Eq. (56)$_4$, a Gibbs energy $G_{RF}(\vartheta_F, c_F^c) = \Psi_{RF}(\vartheta_F, c_F^c) - P_F \vartheta_F$ for the NaCl-water solution is introduced. Based on the definition of the Gibbs energy of the NaCl-water solution, the following equation can be derived:

$$\vartheta_F = -\frac{\partial G_{RF}(P_F, c_F^c)}{\partial P_F}; \quad \bar{\mu}_F^c - \bar{\mu}_F^L = \frac{\partial G_{RF}(\vartheta_F, c_F^c)}{\rho_{RF0} \partial c_F^c} \tag{79}$$

Based on Eq. (79), there is

$$d\vartheta_F = -\frac{\partial^2 G_{RF}(P_F, c_F^c)}{\partial P_F^2} dP_F - \frac{\partial^2 G_{RF}(P_F, c_F^c)}{\partial P_F \partial c_F^c} dc_F^c = \frac{\partial \vartheta_F}{\partial P_F} dP_F - \rho_{RF0} \frac{\partial(\mu_F^c - \mu_F^L)}{\partial P_F} dc_F^c \tag{80}$$

Based on the chemical thermodynamic theory (Annamalai., 2011), assuming that the NaCl-water solution is an ideal liquid solution, the chemical potential can be expressed as:

$$\bar{\mu}_F^c - \bar{\mu}_F^L = \bar{\mu}_F^{c0}(\theta, P_0) - \bar{\mu}_F^{L0}(\theta, P_0) + \frac{R\theta}{M_F^c}\ln[\frac{M_F^L c_F^c}{M_F^c + (M_F^L - M_F^c)c_F^c}] - \frac{R\theta}{M_F^L}\ln[1 - \frac{M_F^L c_F^c}{M_F^c + (M_F^L - M_F^c)c_F^c}] + \bar{V}_F(P_F - P_0)$$ (81)

where $\bar{\mu}_F^{c0}(\theta, P_0)$ and $\bar{\mu}_F^{L0}(\theta, P_0)$ are the standard chemical potential of NaCl solute and water solvent, respectively. $M_F^c$ and $M_F^L$ are the molar mass of NaCl solute and water solvent, respectively. $R$ is the universal gas constant. For a fluid solution, $-\bar{V}_F$ is an approximate constant (Annamalai., 2011). Assuming that the NaCl-water solution is a linear elastic material for the fluid pressure $P_F$. Taking partial derivative of Eq.(81) with respects to $P_F$, and then substituting it into Eq.(80) yields

$$d\vartheta_F = \frac{1}{K_F}dP_F - \rho_{RF0}\bar{V}_F \, dc_F^c$$ (82)

where $K_F$ is the volumetric modulus of the NaCl-water solution. Fig 9 shows the strain variation of NaCl-water solution with the NaCl mass fraction (Simion et al., 2015). Fig 9 indicates that $\bar{V}_F = -0.738/\rho_{RF0}$, where $\rho_{RF0} = 1000 \text{ kg/m}^3$.

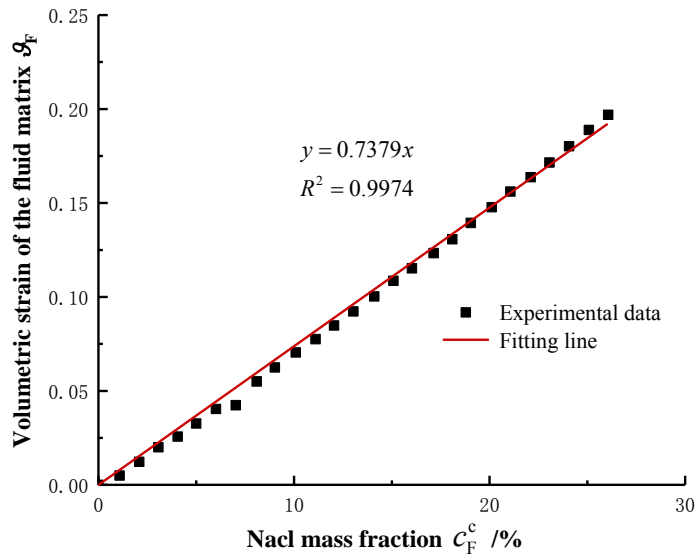

**Fig 9**. The strain variation of the NaCl-water solution with the NaCl mass fraction

Integrating Eq. (82) yields:

$$\vartheta_F = \frac{1}{K_F} P_F - \rho_{RF0} \overline{V}_F c_F^c \qquad (83)$$

Eq. (83) indicates that the strain ($\vartheta_F$) of solution is determined by the pore pressure ($P_F$) and the NaCl mass fraction ($c_F^c$) in the pore NaCl-water solution.

6.6 Darcy's law of the fluid phase

For the confined compression condition, $\Theta_w(W_F)$ can be expressed as $\Theta_w(W_{Fz})$, where $W_{Fz} = v_{Fz} - v_{Sz}$ is the vertical relative velocity of the fluid phase. If $\Theta_w(W_{Fz})$ is adopted as $\gamma_w \varphi_{F0} \varphi_F W_{Fz}^2 / (2k_{Fz})$, where $k_{Fz}$ is the permeability coefficient, $\gamma_w$ is the gravity of water. Substituting the expression of $\Theta_w(W_{Fz})$ into Eq.(65)$_2$ yields:

$$P_F \frac{\partial \varphi_F}{\partial z} - \hat{p}_{Fz} + \hat{p}_{Az} = \frac{\gamma_w \varphi_{F0} \varphi_F}{k_{Fz}} W_{Fz} \qquad (84)$$

The static momentum balance equation for the fluid of NaCl-water solution is

$$\frac{\partial(-\varphi_F P_F)}{\partial z} + \hat{p}_{Fz} = 0 \qquad (85)$$

Substituting Eq.(84) into Eq.(85) by eliminating the $\hat{p}_{Fz}$ of Eq.(85) yields

$$\frac{k_{Fz}}{\gamma_w} \frac{\partial P_F}{\partial z} - \frac{k_{Fz}}{\gamma_w \varphi_F} \hat{p}_{Az} + \varphi_{F0} W_{Fz} = 0 \qquad (86)$$

It is worth noting that under the confined compression condition, only the vertical flow and the vertical deformation are not equal to zero. In this case and the infinitesimal strain condition, the definition of $\hat{p}_{Az}$ becomes:

$$\hat{p}_{Az} = \frac{\partial(\rho_{F0}\xi_F)}{\partial \varepsilon_{Hz}} \frac{\partial \varepsilon_{Hz}}{\partial z} + \frac{\partial(\rho_{F0}\xi_F)}{\partial \varepsilon_{Hz}^p} \frac{\partial \varepsilon_{Hz}^p}{\partial z} \qquad (87)$$

Noting that $\partial(\rho_{F0}\xi_F)/\partial\varepsilon_{Hz} = \partial\Psi_{HF}/\partial\varepsilon_{Hz}$, $\partial(\rho_{F0}\xi_F)/\partial\varepsilon^p_{Hz} = \partial\Psi_{HF}/\partial\varepsilon^p_{Hz}$ and $\Psi_{HF}$ is the function in term of $\varepsilon_{Hz} - \varepsilon^p_{Hz}$. Eq.(87) becomes:

$$\hat{p}_{Az} = \frac{\partial\Psi_{HF}}{\partial\varepsilon_{Hz}} \frac{\partial(\varepsilon_{Hz} - \varepsilon^p_{Hz})}{\partial z} = \frac{\partial\Psi_{HF}(\varepsilon^e_{Hz})}{\partial\varepsilon^e_{Hz}} \frac{\partial\varepsilon^e_{Hz}}{\partial z} \tag{88}$$

Noting that the vertical effective pressure remain constant when the chemical plastic strain of the solid skeleton changes, so the skeleton swelling strain in the elastic model will decrease with the chemical plastic strain. Eqs.(68) ~ (70) and (77) yield:

$$\frac{\partial\Psi_{HF}}{\partial\varepsilon^e_{Hz}} = \tilde{P}_z \left[\exp\frac{-\varepsilon_{Hmax} - \varepsilon^{pc}_{Hz}(c^c_F)}{\lambda^e} - 1\right] \tag{89}$$

Substituting Eqs.(88) and (89) into Eq.(86) yields:

$$\frac{k_{Fz}}{\gamma_w}\frac{\partial P_F}{\partial z} - \frac{k_{Fz}}{\gamma_w \varphi_F}\tilde{P}_z\left[\exp\frac{-\varepsilon_{Hmax} - \varepsilon^{pc}_{Hz}(c^c_F)}{\lambda^e} - 1\right]\frac{\partial\varepsilon^e_{Hz}}{\partial z} + \varphi_{F0}W_{Fz} = 0 \tag{90}$$

For non-swelling material, $\varepsilon_{Hmax} + \varepsilon^{pc}_{Hz}(c^c_F) = 0$, Eq. (90) is degenerated into the classical Darcy law. According to the experiment data, the permeability coefficient of bentonite will change with the skeleton volumetric strain and NaCl mass fraction (Castellanos, et al, 2008).

Noting that $\hat{c}_F$ is assumed to be zero in the current study of bentonite. Based on $\rho_{F0} = \varphi_{F0}\rho_{RF0}$, $\rho_F = \varphi_F\rho_{RF}$ and Eq.(12)$_2$, we have under the infinitesimal strain case:

$$\varepsilon_{FV} = \frac{\rho_{RF0} - \rho_{RF}}{\rho_{RF0}} + \frac{\varphi_{F0} - \varphi_F}{\varphi_{F0}} \tag{91}$$

where $\varepsilon_{FV}$ is the volumetric strain of the fluid. Using $\varphi_{S0} + \varphi_{F0} = 1$ and $\varphi_S + \varphi_F = 1$ and Eqs.(50) and (51) yields (Hu, 2016)

$$\varepsilon_{FV} = -\vartheta_F - \frac{\varphi_{S0}}{\varphi_{F0}}\varepsilon_{HV} \tag{92}$$

where $\varepsilon_{HV}$ is the solid skeleton volumetric strain. In practical engineering, it is more concerned about the fluid flowing out or into the saturated porous media. Biot's increment of fluid content (fluid volume accumulation per unit bulk volume) is defined as $\zeta_F = \varphi_{F0}(\varepsilon_{FV} - \varepsilon_{SV})$. Using Eqs. (53) and (92) yields:

$$\zeta_F = \varphi_{F0}(\varepsilon_{FV} - \varepsilon_{SV}) = -\varepsilon_{HV} + \varphi_{F0}(\vartheta_S - \vartheta_F) \tag{93}$$

For the confined compression condition, noting that (i) only the vertical flow and the vertical deformation are not equal to zero, (ii) as the above-mentioned, it is assumed that only the solid matrix and skeleton vertical strains is not equal to zero in order to simplify the analysis. Under these conditions, the volumetric strain is equal to the vertical strain, i.e. $\varepsilon_{SV} = \varepsilon_{Sz}, \varepsilon_{HV} = \varepsilon_{Hz}$, $\vartheta_S = \vartheta_{Sz}$ and $\zeta_F = \zeta_{Fz}$, After using Eq.(63), Eq.(93) becomes:

$$\zeta_{Fz} = -\varepsilon_{Hz} + \varphi_{F0}(\vartheta_{Sz} - \vartheta_F) = -\varepsilon_{Sz} - \varphi_{S0}\vartheta_{Sz} - \varphi_{F0}\vartheta_F \tag{94}$$

Differentiating Eq.(94) with respect to time ($t$) yielding:

$$\frac{d\zeta_{Fz}}{dt} = -\frac{d\varepsilon_{Sz}}{dt} - \varphi_{S0}\frac{d\vartheta_{Sz}}{dt} - \varphi_{F0}\frac{d\vartheta_F}{dt} \tag{95}$$

Substituting $\varphi_{F0} + \varphi_{S0} = 1$ and Eqs. (67), (78) and (83) into Eq.(95) yields:

$$\frac{d\zeta_{Fz}}{dt} = \frac{\lambda}{\tilde{P}_z}\frac{d\tilde{P}_z}{dt} + \varepsilon_P^c \lambda^c \exp(-\lambda^c c_F^c)\frac{dc_F^c}{dt} + \frac{\varphi_{F0}}{E_{RS}}\frac{dP_S}{dt} - \frac{\varphi_{F0}}{K_F}\frac{dP_F}{dt} + \rho_{F0}\bar{V}_F\frac{dc_F^c}{dt} \tag{96}$$

Using equation $\zeta_F = \varphi_{F0}(\varepsilon_{FV} - \varepsilon_{SV})$ and making the divergence of Eq.(90) with respect to z yields:

$$\frac{\partial}{\partial z}\left\{\frac{k_{Fz}}{\gamma_w}\frac{\partial P_F}{\partial z} - \frac{k_{Fz}}{\gamma_w \varphi_F}\tilde{P}_z\left[\exp\frac{-\varepsilon_{Hmax} - \varepsilon_{Hz}^{pc}(c_F^c)}{\lambda^e} - 1\right]\frac{\partial \varepsilon_{Hz}^e}{\partial z}\right\} = -\frac{\partial}{\partial z}(\varphi_{F0}W_{Fz}) = -\frac{\partial \zeta_{Fz}}{\partial t} \tag{97}$$

Incorporating Eqs. (96) with (97) gives:

$$\frac{\lambda}{\tilde{P}_z}\frac{d\tilde{P}_z}{dt} + \varepsilon_p^c \lambda^c \exp(-\lambda^c c_F^c)\frac{dc_F^c}{dt} + \frac{\varphi_{F0}}{E_{RS}}\frac{dP_S}{dt} - \frac{\varphi_{F0}}{K_F}\frac{dP_F}{dt} + \rho_{F0}\overline{V}_F \frac{dc_F^c}{dt} =$$
$$-\frac{\partial}{\partial z}\left\{\frac{k_{Fz}}{\gamma_w}\frac{\partial P_F}{\partial z} - \frac{k_{Fz}}{\gamma_w \varphi_F}\tilde{P}_z \left[\exp\frac{-\varepsilon_{Hmax} - \varepsilon_{Hz}^{pc}(c_F^c)}{\lambda^e} - 1\right]\frac{\partial \varepsilon_{Hz}^e}{\partial z}\right\} \quad (98)$$

Eq. (98) is the consolidation governing equation of bentonite considering the swelling behavior and the chemical activity.

6.7 Seepage-diffusion equation of NaCl solute

For the confined compression condition, only the vertical flow and the vertical deformation are not equal to zero. In this case, $b_F^\beta$ is degenerated into $b_{Fz}^\beta$, and $u_F^c$ is degenerated into $u_{Fz}^c$. Assuming that $\Theta_d(\rho_F^c u_{Fz}^c) = k_d(\rho_F^c u_{Fz}^c)^2/2$, Eq. (65c) yields:

$$\frac{\partial(\overline{\mu}_F^c - \overline{\mu}_F^L)}{\partial z} - b_{Fz}^c + b_{Fz}^L = -\frac{\partial \Theta_d(\rho_F^c u_{Fz}^c)}{\partial(\rho_F^c u_{Fz}^c)} = -k_d \rho_F^c u_{Fz}^c \quad (99)$$

where $k_d$ is the diffusion coefficient. Noting that only the vertical flow occurs in the confined compression condition and $\hat{c}_F^c - c_F^c \hat{c}_F$ is assumed to be zero in the current study of bentonite, substituting Eq. (99) into Eq.(16)$_2$ and using $v_{Fz}^c = u_{Fz}^c + v_{Fz} = u_{Fz}^c + W_{Fz} + v_{Sz}$ and Eqs. (9) and (99), Eq. (16)$_2$ yields:

$$\frac{\partial c_F^c}{\partial t} + W_{Fz}\frac{\partial c_F^c}{\partial z} - \frac{1}{k_d \rho_F^c}\left[\frac{\partial(\overline{\mu}_F^c - \overline{\mu}_F^L)}{\partial z} - b_{Fz}^c + b_{Fz}^L\right]\frac{\partial c_F^c}{\partial z} + v_{Sz}\frac{\partial c_F^c}{\partial z} -$$
$$\frac{1}{\rho_F}\frac{\partial}{\partial z}\left[\frac{\partial(\overline{\mu}_F^c - \overline{\mu}_F^L)}{k_d \partial z} - \frac{b_{Fz}^c - b_{Fz}^L}{k_d}\right] = 0 \quad (100)$$

Under the infinitesimal condition, $v_{Sz} \approx 0$, $\rho_F \approx \rho_{F0}$ and $\rho_F^c \approx \rho_{F0}^c$. Using these equations, Eq. (100) can be degenerated into:

$$\frac{\partial c_F^c}{\partial t} + W_{Fz}\frac{\partial c_F^c}{\partial z} - \frac{1}{k_d \rho_{F0}^c}\left[\frac{\partial(\overline{\mu}_F^c - \overline{\mu}_F^L)}{\partial z} - b_{Fz}^c + b_{Fz}^L\right]\frac{\partial c_F^c}{\partial z} -$$
$$\frac{1}{\rho_{F0}}\frac{\partial}{\partial z}\left[\frac{\partial(\overline{\mu}_F^c - \overline{\mu}_F^L)}{k_d \partial z} - \frac{b_{Fz}^c - b_{Fz}^L}{k_d}\right] = 0 \quad (101)$$

The partial molar volume $\bar{V}_{m,L}$ of water solvent is an approximate constant, so for an ideal NaCl-water solution, its chemical potential of the water solvent can be expressed as：

$$\bar{\mu}_F^L = \bar{\mu}_F^{L0}(\theta, P_0) + \frac{R\theta}{M_F^L}\ln[1 - \frac{M_F^L c_F^c}{M_F^c + (M_F^L - M_F^c)c_F^c}] \\ + \frac{\bar{V}_{m,L}}{M_F^L}(P_F - P_0) \tag{102}$$

where $\bar{V}_{m,L}$ is the partial molar volume of water solvent. Substituting Eq. (81) into Eq. (102) gives

$$\bar{\mu}_F^c = \bar{\mu}_F^{c0}(\theta, P_0) + \frac{R\theta}{M_F^c}\ln[\frac{M_F^L c_F^c}{M_F^c + (M_F^L - M_F^c)c_F^c}] + \\ \frac{\bar{V}_{m,L} + M_F^L \bar{V}_F}{M_F^L}(P_F - P_0) \tag{103}$$

Eqs. (102) and (103) yield:

$$\bar{\mu}_F^c - \bar{\mu}_F^L = \bar{\mu}_F^{c0}(\theta, P_0) - \bar{\mu}_F^{L0}(\theta, P_0) + \frac{R\theta}{M_F^c}\ln[\frac{M_F^L c_F^c}{M_F^c + (M_F^L - M_F^c)c_F^c}] - \\ \frac{R\theta}{M_F^L}\ln[1 - \frac{M_F^L c_F^c}{M_F^c + (M_F^L - M_F^c)c_F^c}] + \bar{V}_F(P_F - P_0) \tag{104}$$

Substituting Eq. (104) into Eq. (101) yields:

$$\frac{\partial c_F^c}{\partial t} + W_{Fz}\frac{\partial c_F^c}{\partial z} - \frac{1}{k_d \rho_{F0}^c}[\frac{R\theta / c_F^c /(1-c_F^c)}{M_F^c + (M_F^L - M_F^c)c_F^c}\frac{\partial c_F^c}{\partial z} + \bar{V}_F\frac{\partial P_F}{\partial z} \\ -b_{Fz}^c + b_{Fz}^L]\frac{\partial c_F^c}{\partial z} - \frac{1}{\rho_{F0}}\frac{\partial}{\partial z}[\frac{R\theta / c_F^c /(1-c_F^c)}{k_d[M_F^c + (M_F^L - M_F^c)c_F^c]}\frac{\partial c_F^c}{\partial z} \\ + \frac{\bar{V}_F}{k_d}\frac{\partial P_F}{\partial z} - \frac{b_{Fz}^c - b_{Fz}^L}{k_d}] = 0 \tag{105}$$

Eq. (105) is the seepage-diffusion law of NaCl solute in the bentonite. When $W_{Fz} = 0$, $\partial c_F^c / \partial t = 0$ only if the following equation is hold:

$$\frac{R\theta / c_F^c /(1-c_F^c)}{M_F^c + (M_F^L - M_F^c)c_F^c}\frac{\partial c_F^c}{\partial z} + \bar{V}_F\frac{\partial P_F}{\partial z} - b_{Fz}^c + b_{Fz}^L = 0 \tag{106}$$

The action of semipermeable membrane on the NaCl solute diffusion is equivalent to the equilibrium of the chemical potential of NaCL solute under the external load, so $b_{Fz}^c = \partial \mu_F^c / \partial z$. Substituting $b_{Fz}^c = \partial \mu_F^c / \partial z$ and Eq. (103) into (106) yields:

$$\frac{R\theta}{M_F^c - M_F^L c_F^c} \frac{\partial c_F^c}{\partial z} - \frac{\overline{V}_{m,L}}{M_F^L} \frac{\partial P_F}{\partial z} + b_{Fz}^L = 0 \tag{107}$$

Integrating Eq.(107) yields that the equilibrium expression of the water solvent chemical potential is:

$$P_F - \int_0^z \frac{M_F^L}{\overline{V}_{m,L}} b_{Fz}^L \, dz = -\frac{R\theta}{\overline{V}_{m,L}} \ln[1 - \frac{M_F^L c_F^c}{M_F^c + (M_F^L - M_F^c)c_F^c}] + C \tag{108}$$

where $C$ is constant. Noting that $P_F - \int_0^z (M_F^L b_{Fz}^L / \overline{V}_{m,L}) \, dz$ is zero when $c_F^c = 0$, so $C = 0$. Since $c_F^c$ is a very small quantity, we have:

$$\ln[1 - \frac{M_F^L c_F^c}{M_F^c + (M_F^L - M_F^c)c_F^c}] \approx -\frac{M_F^L c_F^c}{M_F^c + (M_F^L - M_F^c)c_F^c} \tag{109}$$

Substituting $C = 0$ and Eq. (109) into Eq.(108) yields:

$$\begin{aligned} P_F - \int_0^z \frac{M_F^L}{\overline{V}_{m,L}} b_{Fz}^L \, dz &= \frac{R\theta}{\overline{V}_{m,L}} \frac{M_F^L c_F^c}{M_F^c + (M_F^L - M_F^c)c_F^c} \\ &= \frac{R\theta}{\overline{V}_{m,L}} \frac{M_F^L \rho_F^c}{M_F^c \rho_F^L + M_F^L \rho_F^c} = \frac{R\theta}{\overline{V}_{m,L}[(\rho_F^c / M_F^c) + (\rho_F^L / M_F^L)]} \frac{\rho_F^c}{M_F^c} \end{aligned} \tag{110}$$

The right side of the last equation in Eq. (110) is the same as the expression of the classical osmotic pressure (Johnson, et al., 2021), so $P_F - \int_0^z (M_F^L b_{Fz}^L / \overline{V}_{m,L}) \, dz$ is the classical osmotic pressure. The derivation of Eqs. (105)~(110) show the classical osmotic pressure can be derived from the seepage-diffusion equation of NaCl solute (i.e. Eq.(105)) and the classical osmotic pressure hold only under a very special condition ( $W_{Fz} = 0$ and $\partial c_F^c / \partial t = 0$ ) in the saturated bentonite. $W_{Fz} = 0$ and

$\partial c_F^c / \partial t = 0$ mean both the momentum of the fluid phase and the chemical potential of the constituents in the fluid phase reach the thermodynamic balance. But there exists the consolidation process of bentonite after the external action is loaded, in this case $W_{Fz} \neq 0$, so both $W_{Fz} = 0$ and $\partial c_F^c / \partial t = 0$ do not always hold for the saturated porous media. Therefore, for the saturated porous media, the expression of the classical osmotic pressure is not hold at any case.

**7. Conclusions**

(1) The solid deformation can be decomposed into the porosity-dependent skeleton deformation, the solid matrix deformation related to the material deformation, and the mass-exchange deformation related to the mass supply. The porosity-dependent skeleton strain is isolated from the solid and fluid strains in order to highlight the important role of porosity change on the hydro-mechanical-chemical multi-field coupling effect.

(2) The energy equation of saturated porous media is expressed in terms of energy-conjugation based on the HMT and the chemical thermodynamics. A constitutive theoretical framework of saturated porous media considering chemical activity has been established. The studies show that the elastic constitutive equations and the combination relationship between the elastic and plastic mechanic element can be derived from the free energy potential function. The irreversible constitutive equation such as the plastic equation and the constitutive model between thermodynamic flux and force can be derived from the dissipative potential function.

(3) Under the guidance of the above constitutive theoretical framework, a one-dimensional constitutive model considering the hydro-mechanical-chemical coupling is proposed for the bentonite saturated with NaCl-Water solution based on the confine compression experiments. This coupling constitutive model includes the constitutive equation of the solid phase, the constitutive equation of fluid phase and the seepage-diffusion equation of salt solute, in which the interactions among the chemical activity, the solid skeleton deformation, the matrix deformation and the seepage-diffusion of NaCl solute are carefully studied.

**Reference**

Achanta, S., Okos, M. R., Cushman, J. H., & Kessler, D. P. (1997). Moisture transport in shrinking gels during saturated drying. AIChE Journal, 43(8), 2112-2122. https://doi.org/10.1002/aic.690430818

Alonso, E. E., Gens, A., & Josa, A. (1990). A constitutive model for partially saturated soils. Géotechnique, 40(3), 405-430. https://doi:10.1680/geot.1990. 40.3.405

Annamalai, K., Puri, I. K., & Jog, M. A. (2011). Advanced thermodynamics engineering. CRC press. https://doi.org/10.1201/9781439805718

Bai, B., Zhang, J., Liu, L., & Ji, Y. (2020). The deposition characteristics of coupled lead ions and suspended silicon powders along the migration distance in water seepage. Transport in Porous Media, 134(3), 707-724. https://doi.org/10.1007/s11242-020-01464-3

Bennethum, L. S., Cushman, J. H., & Murad, M. A. (1996). Clarifying mixture theory

and the macroscale chemical potential for porous media. International Journal of Engineering Science, 34(14), 1611-1621. https://doi.org/10.1016/S0020-7225(96)00042-0

Bennethum, L. S., Murad, M. A., & Cushman, J. H. (2000). Macroscale thermodynamics and the chemical potential for swelling porous media. Transport in Porous Media, 39(2), 187-225. https://doi.org/10.1023/A:1006661330427

Biot, M. A., and Willis, D. G., (1957). The elastic coefficient of the theory of consolidation. J.Appl. Mech.oe, 5: 94-601. https://doi.org/10.1115/1.4011606

Bowen, R. M. (1976). Theory of Mixtures, vol. III. Continuum Physics: Mixtures and EM Field Theories, 1-127.

Bowen, R. M. (1982). Compressible porous media models by use of the theory of mixtures. International Journal of Engineering Science, 20(6), 697-735. https://doi.org/10.1016/0020-7225(82)90082-9

Borja, R. I. (2006). On the mechanical energy and effective stress in saturated and unsaturated porous continua. International Journal of Solids and Structures, 43(6), 1764-1786. https://doi.org/10.1016/j.ijsolstr.2005.04.045

Carroll, M.M., Katsube, N., (1983). The role of Terzaghi effective stress in linearly elastic deformation. J. Energy Resour. Technol. 105, 509-511.

Castellanos, E., Villar, M. V., Romero, E., Lloret, A., & Gens, A. (2008). Chemical impact on the hydro-mechanical behaviour of high-density FEBEX bentonite. Physics and Chemistry of the Earth, Parts A/B/C, 33, S516-S526. https://doi.org/

10.1016/j.pce.2008.10.056

Chen, Y., Ke, H., Fredlund, D. G., Zhan, L., & Xie, Y. (2010). Secondary compression of municipal solid wastes and a compression model for predicting settlement of municipal solid waste landfills. Journal of geotechnical and geoenvironmental engineering, 136(5), 706-717. https://doi.org/10.1061/(ASCE)GT.1943-5606.0000273

Chen, X., Pao, W., Thornton, S., & Small, J. (2016). Unsaturated hydro-mechanical-chemical constitutive coupled model based on mixture coupling theory: hydration swelling and chemical osmosis. International Journal of Engineering Science, 104, 97-109. https://doi.org/10.1016/j.ijengsci.2016.04.010

Cheng, A.H.-D., (2016). Poroelasticity, vol. 27. Springer, Berlin.

Chapuis, R. P. (2004). Predicting the saturated hydraulic conductivity of sand and gravel using effective diameter and void ratio. Canadian geotechnical journal, 41(5), 787-795. https://doi.org/10.1139/t04-022

Collins, I.F., Houlsby, G.T., (1997). Application of thermo-mechanical principles to the modeling of geotechnical materials. Proceedings of the Royal Society of London A, 45(3):1975- 2001. https://doi.org/10.1098/rspa.1997.0107

Coussy, O., (2004). Poromechanics. John Wiley & Sons, Chichester

Cushman, J. H., Bennethum, L. S., & Hu, B. X. (2002). A primer on upscaling tools for porous media. Advances in Water Resources, 25(8-12), 1043-1067. https://doi.org/10.1016/S0309-1708(02)00047-7

De Groot, S. R., & Mazur, P. (1962). Non-equilibrium thermodynamics. McGraw-Hill, New York.

Detmann, B. (2021). Modeling chemical reactions in porous media: a review. Continuum Mechanics and Thermodynamics, 33(6), 2279-2300. https://doi.org/10.1007/s00161-021-01049-5

Detournay, E., Cheng, A.H.-D., (1993). Fundamentals of poroelasticity. In: Analysis and Design Methods. Elsevier, pp. 113-171.

Di Maio, C., Santoli, L. and Schiavone, P., (2004). Volume change behaviour of clays: the influence of mineral composition, pore fluid composition and stress state. Mechanics of Materials, 36(5-6): 435-451. doi:10.1016/S0167-6636(03)00070-X

Dehghani, H., Penta, R., & Merodio, J. (2018). The role of porosity and solid matrix compressibility on the mechanical behavior of poroelastic tissues. Materials Research Express, 6(3), 035404. https://orcid.org/0000-0003-1202-8775

Dominijanni, A., Manassero, M., & Puma, S. (2014). Coupled chemical-hydraulic-mechanical behaviour of bentonites. In Bio-and Chemo-Mechanical Processes in Geotechnical Engineering: Géotechnique Symposium in Print 2013 (pp. 57-71). ICE Publishing. https://doi.org/10.1680/bcmpge.60531.005

Gajo, A., Loret, B., & Hueckel, T. (2002). Electro-chemo-mechanical couplings in saturated porous media: elastic–plastic behaviour of heteroionic expansive clays.

International journal of solids and structures, 39(16), 4327-4362. https://doi.org/10.1016/S0020-7683(02)00231-7

Geertsma, J. (1957). The effect of fluid pressure decline on volumetric changes of porous rocks. Transactions of the AIME, 210(01), 331-340. https://doi.org/10.2118/728-G

Guimarães, L. D. N., Gens, A., Sánchez, M., & Olivella, S. 2014. A chemo-mechanical constitutive model accounting for cation exchange in expansive clays. In Bio-and Chemo-Mechanical Processes in Geotechnical Engineering: Geotechnique Symposium in Print 2013 (pp. 18-31). ICE Publishing. https://doi.org/10.1680/bcmpge.60531.002

Hassanizadeh, M., & Gray, W. G. (1979a). General conservation equations for multi-phase systems: 1. Averaging procedure. Advances in water resources, 2, 131-144. https://doi.org/10.1016/0309-1708(79)90025-3

Hassanizadeh, M., & Gray, W. G. (1979b). General conservation equations for multi-phase systems: 2. Mass, momenta, energy, and entropy equations. Advances in water resources, 2, 191-203. https://doi.org/10.1016/0309-1708(79)90035-6

Houlsby, G. T. (1997). The work input to an unsaturated granular material. Géotechnique, 47(1), 193-196. https://doi.org/10.1680/geot.1997.47.1.193

Hu, Y.Y, (2008) On Application of Non-equilibrium thermodynamics in Environmental Geotechnical Project. Journal of Lishui University, 5.

Hu, Y.Y, (2016). Study on the super viscoelastic constitutive theory for saturated po-rous media. Applied Mathematics and Mechanics, 37(6), 584-598. (in Chinese) doi: 10.3879/j.issn.1000-0887.2016.06.004

Johnson, D., Hashaikeh, R., & Hilal, N. (2021). Basic principles of osmosis and osmotic pressure. In Osmosis Engineering (pp. 1-15). https://doi.org/10.1016/B978-0-12-821016-1.00011-5

Lambe, T. W., & Whitman, R. V. (1991). Soil mechanics (Vol. 10). John Wiley & Sons.

Li, H, (2021). Degradation-Consolidation Behaviors and Treatment Methods for the Liquid-Gas induced Environmental Disaster of High-Food-Waste-Content MSW Landfill (in Chinese).

Loret, B., Hueckel, T., & Gajo, A. (2002). Chemo-mechanical coupling in saturated porous media: elastic–plastic behaviour of homoionic expansive clays. International Journal of Solids and Structures, 39(10), 2773-2806. https://doi.org/10.1016/S0020-7683(02)00151-8

Ma, T., Wei, C., Chen, P., & Li, W. (2019). Chemo-mechanical coupling constitutive model for chalk considering chalk–fluid physicochemical interaction. Géotechnique, 69(4), 308-319. https://doi.org/10.1680/jgeot.17.P.115

Ma, Y., Chen, X. H., Hosking, L. J., Yu, H. S., & Thomas, H. R. (2022). THMC constitutive model for membrane geomaterials based on Mixture Coupling Theory. International Journal of Engineering Science, 171, 103605. https://doi.org/10.1016/j.ijengsci.2021.103605

Santamarina, J. C., Klein, K. A., Palomino, A., & Guimaraes, M. S. (2018). Micro-scale aspects of chemical-mechanical coupling: Interparticle forces and fabric. In Chemo-mechanical coupling in clays (pp. 47-58). Routledge. https://doi.org/ 10.1201/9781315139289

Simion, A. I., Grigoraş, C. G., Roşu, A. M., & Gavrilă, L. (2015). Mathematical modelling of density and viscosity of NaCl aqueous solutions. Journal of Agroalimentary Processes and Technologies, 21(1), 41-52.

Thomas, H. R., Sedighi, M., & Vardon, P. J. (2012). Diffusive reactive transport of multicomponent chemicals under coupled thermal, hydraulic, chemical and mechanical conditions. Geotechnical and Geological Engineering, 30(4), 841-857. https://doi.org/10.1007/s10706-012-9502-9

Thomas, H.R., Vardon, P.J., Li, YC. (2010). Coupled Thermo-Hydro-Chemo-Mechanical Modeling for Geoenvironmental Phenomena. In: Chen, Y., Zhan, L., Tang, X. (eds) Advances in Environmental Geotechnics. Springer, Berlin, Heidelberg. https://doi.org/10.1007/978-3-642-04460-1_24

Wei, C. (2014). A theoretical framework for modeling the chemo-mechanical behavior of unsaturated soils. Vadose Zone Journal, 13(9), 1-21. https://doi.org/10.2136/vzj2013.07.0132

Xie, H., Thomas, H. R., Chen, Y., Sedighi, M., Zhan, T. L., & Tang, X. (2015). Diffusion of organic contaminants in triple-layer composite liners: an analytical modeling approach. Acta Geotechnica, 10(2), 255-262.

https://doi.org/10.1007/s11440-013-0262-3

Xu, Y. (2019). Modified effective stress induced by osmotic suction and its validation in volume change and shear strength of bentonite in saline solutions. Chinese Journal of Geotechnical Engineering, 41(4), 631-638. (in Chinese) https://doi.org/10.11779/cjge201904005

Yang, D., Yan, R., Ma, T., & Wei, C. (2022). Compressive behavior of kaolinitic clay under chemo-mechanical loadings. Acta Geotechnica, 1-18. https://doi.org/10.1007/s11440-022-01554-0

Ye, W. M., Zhang, F., Chen, B., Chen, Y. G., Wang, Q., & Cui, Y. J. (2014). Effects of salt solutions on the hydro-mechanical behavior of compacted GMZ01 Bentonite. Environmental earth sciences, 72(7), 2621-2630. doi: 10.1007/s12665-014-3169-x

Zhang, F., Ye, W. M., Chen, Y. G., Chen, B., & Cui, Y. J. (2016). Influences of salt solution concentration and vertical stress during saturation on the volume change behavior of compacted GMZ01 bentonite. Engineering Geology, 207, 48-55. https://doi.org/10.1016/j.enggeo.2016.04.010

Zhang, F., Ye, W. M., Wang, Q., Chen, Y. G., & Chen, B. (2020). Effective stress incorporating osmotic suction and volume change behavior of compacted GMZ01 bentonite. Acta Geotechnica, 15(7), 1925-1934. https://doi.org/10.1007/s11440-019-00906-7

Zhang, Z., & Cheng, X. (2015). A thermodynamic constitutive model for undrained monotonic and cyclic shear behavior of saturated soils. Acta Geotechnica, 10(6), 781-796. https://doi.org/10.1007/s11440-015-0389-5

Zhang, Z. (2017). A thermodynamics-based theory for thethermo-poro-mechanical modeling of saturated clay. International Journal of Plasticity, 92, 164-185. https://doi.org/10.1016/j.ijplas.2017.03.007

**Declaration of Competing Interest**

The authors declare that they have no known competing financial interests or personal relationships that could have appeared to influence the work reported in this paper.

**Acknowledgments**

This research work was financially supported by the National Natural Science Foundation of China, Grant No. 52178360.